\documentclass[arguments]{aastex631}

\usepackage{amsmath}
\pdfoutput=1
\definecolor{ForestGreen}{RGB}{34,139,34}
\newcommand{\mycirc}[1][black]{\normalsize{\textcolor{#1}{\ensuremath\bullet}}\normalsize}
\shorttitle{}
\shortauthors{Sai Swagat et al.}
\graphicspath{{./}{figures/}}
\begin{document}

\title{Constraining extended teleparallel gravity via cosmography: A model-independent approach}

\author[0000-0003-0580-0798]{Sai Swagat Mishra}
\affiliation{Department of Mathematics, Birla Institute of Technology and
Science-Pilani,\\ Hyderabad Campus, Hyderabad-500078, India.}

\author[0000-0001-8561-130X]{N. S. Kavya}
\affiliation{Department of P.G. Studies and Research in Mathematics,
 \\
 Kuvempu University, Shankaraghatta, Shivamogga 577451, Karnataka, INDIA}

%\collaboration{6}{(AAS Journals Data Editors)}

\author[0000-0003-2130-8832]{P.K. Sahoo}\thanks{pksahoo@hyderabad.bits-pilani.ac.in}
\affiliation{Department of Mathematics, Birla Institute of Technology and
Science-Pilani,\\ Hyderabad Campus, Hyderabad-500078, India.}
%\affiliation{AAS Journals Associate Editor-in-Chief}

\author[0000-0002-2799-2535]{V. Venkatesha}
\affiliation{Department of P.G. Studies and Research in Mathematics,
 \\
 Kuvempu University, Shankaraghatta, Shivamogga 577451, Karnataka, INDIA}

%% Note that the \and command from previous versions of AASTeX is now
%% depreciated in this version as it is no longer necessary. AASTeX 
%% automatically takes care of all commas and "and"s between authors names.

%% AASTeX 6.31 has the new \collaboration and \nocollaboration commands to
%% provide the collaboration status of a group of authors. These commands 
%% can be used either before or after the list of corresponding authors. The
%% argument for \collaboration is the collaboration identifier. Authors are
%% encouraged to surround collaboration identifiers with ()s. The 
%% \nocollaboration command takes no argument and exists to indicate that
%% the nearby authors are not part of surrounding collaborations.

%% Mark off the abstract in the ``abstract'' environment. 
\begin{abstract}

 As a classical approach, the dynamics of the Universe, influenced by its dark components, are unveiled through prior modifications of Einstein's equations. Cosmography, on the other hand, is a highly efficient tool for reconstructing any modified theory in a model-independent manner. By employing kinematic variables, it offers a profound explanation for cosmic expansion. Although the cosmographical approach has been highly successful in several geometric theories in recent years, it has not been extensively explored in coupled gravities. With this in mind, we intend to constrain an extended teleparallel gravity model, $f(T,\mathcal{T})$, through cosmographic parameters. We utilize Taylor series expansion, assuming a minimally coupled form, to constrain the unknowns involved in the series. To achieve this, we conduct a Markov Chain Monte Carlo analysis (MCMC) using three different datasets (CC, BAO, and Pantheon+SH0ES). The constrained results obtained from MCMC are then compared and verified using various cosmological parameters. Finally, we compare the resulting models with \textbf{three} well-known $f(T,\mathcal{T})$ models.

\end{abstract}

%% Keywords should appear after the \end{abstract} command. 
%% The AAS Journals now uses Unified Astronomy Thesaurus concepts:
%% https://astrothesaurus.org
%% You will be asked to selected these concepts during the submission process
%% but this old "keyword" functionality is maintained in case authors want
%% to include these concepts in their preprints.

%% From the front matter, we move on to the body of the paper.
%% Sections are demarcated by \section and \subsection, respectively.
%% Observe the use of the LaTeX \label
%% command after the \subsection to give a symbolic KEY to the
%% subsection for cross-referencing in a \ref command.
%% You can use LaTeX's \ref and \label commands to keep track of
%% cross-references to sections, equations, tables, and figures.
%% That way, if you change the order of any elements, LaTeX will
%% automatically renumber them.
%%
%% We recommend that authors also use the natbib \citep
%% and \citet commands to identify citations.  The citations are
%% tied to the reference list via symbolic KEYs. The KEY corresponds
%% to the KEY in the \bibitem in the reference list below. 

\section{Initiation} \label{sec:intro}

In modern cosmology, a significant challenge lies in precisely understanding the influence of components of the Universe on its dynamics during expansion, particularly in the light of the confirmed acceleration as demonstrated in \cite{SupernovaCosmologyProject:1997czu,SupernovaSearchTeam:1998fmf}. While the concordance model is well-supported by observations, it falls short in addressing the coincidence problem \cite{Zlatev:1998tr} concerning the vastly differing magnitudes of dark matter (DM) and dark energy (DE), as well as the fine-tuning discrepancy between the predictions from quantum gravity and the current observational constraints on the value of $\Lambda$ \cite{Steinhardt:1999nw, Abdalla:2022yfr,Weinberg:1988cp,Tsujikawa:2010sc,Burgess:2013ara}. Therefore, an alternative explanation could be that the substance responsible for initiating the present acceleration of the universe might not be attributed to $\Lambda$. This necessitates the exploration of models beyond $\Lambda$CDM. Though quintessence models, featuring scalar fields with potential, offer a viable explanation for cosmological data with some degree of adjustment, the lack of strongly motivated scalar field models supported by theoretical particle physics remains notable. Thus, one could consider that the effect of cosmic acceleration is perhaps explained within the framework of modified gravities, where extensions of General Relativity (GR) are taken into account. 

Here, we examine an extension of gravitational theory referred to as the Teleparallel Equivalent of General Relativity (TEGR), as originally proposed by Einstein himself in \cite{Unzicker:2005in,Maluf:2002zc,Obukhov:2002tm,Ferraro:2006jd,Linder:2010py}, wherein the gravitational field is defined by the torsion tensor. Although most of the results of TEGR coincide, the extensions of TEGR need not necessarily coincide with corresponding extensions of GR \cite{Cai:2015emx}. Indeed, these formulations lead to different outcomes. Thus, in recent years, torsion-based modified theories have been explored extensively in a wide range \cite{Wu:2010mn,Chen:2010va,Dent:2010nbw,Bahamonde:2021srr,Moreira:2021xfe,Escamilla-Rivera:2019ulu,Kofinas:2014daa,Gonzalez-Espinoza:2021mwr}. In \cite{Harko:2014aja}, Harko et al. presented a new gravitational theory known as $f(T,\mathcal{T})$, in which the Lagrangian of geometry is constructed with a coupling between the torsion element and the trace of the energy-momentum tensor (for some of the prominent works in this theory, readers can refer to the references \cite{Mishra:2023khd,Ghorui:2023ssn,Mishra:2023onl,Mandal:2023bzo,Farrugia:2016pjh,Junior:2015bva} therein).

In this paper, we explore a model-independent method for reconstructing the $f(T,\mathcal{T})$ function based on cosmological observations. The reconstruction of a gravitational theory involves determining the functional form of the gravity model based on certain criteria. For instance, references \cite{Gadbail:2024syv,Vogt:2024pws,Gomes:2023xzk,Panda:2023jwe,Saez-Gomez:2016wxb,Chen:2024xkv} utilized diverse methods to reconstruct gravitational theories. Among these, some rely on a particular gravity model, while others utilize model-independent techniques such as the application of Raychaudhuri equations \cite{Gadbail:2024syv,Panda:2023jwe,Choudhury:2019zod} and observational data \cite{Vogt:2024pws,Gomes:2023xzk}. In our article, we intend to explore a notable model-independent approach known as \textit{Cosmography}. It presents a broad approach for handling cosmological parameters using kinematic quantities. We can utilize these quantities to examine the dynamics of the universe without assuming a specific cosmological model.

In the literature, numerous studies have explored cosmography within geometry theories such as $f(R)$, $f(T)$, and $f(Q)$ \cite{Capozziello:2008qc,Capozziello:2011hj,Sabiee:2022iyo,Mandal:2020buf}. However, there has been relatively less emphasis on reconstructing coupled gravities. Thus, in this work, we consider a generic family of minimal functional forms of $f(T,\mathcal{T})$ gravity and deduce an observationally agreeable scenario for an arbitrary function. The kinematical cosmographic series approach involves a greater number of parameters that need to be constrained by data compared to the simplest cosmological models. To streamline the analysis, we assume flat spacetime and uphold the validity of the cosmological principle. These minimal assumptions help simplify the problem, enabling us to employ kinematical cosmography as a model-independent tool to ascertain the compatibility of various $f(T,\mathcal{T})$ models with current observations.

The outline of this paper is as follows: \autoref{sec:II} provides comprehensive details regarding the geometric foundation of $f(T,\mathcal{T})$ gravity. In \autoref{sec:III}, one can see the cosmographic analysis and governing field equations of isotropic and homogeneous universe. In \autoref{sec:IV}, we present the observational constraints, including a cosmographic description of parameters, datasets for MCMC analysis, statistical methodology, and the corresponding results. In the final \autoref{sec:V}, we conclude with a detailed interpretation and discussion of the results. 

\section{Introduction to $f(T,\mathcal{T})$, The Extended Teleparallel Theory}\label{sec:II}
In recent years, torsion theories have been very successful in describing late-time scenarios. Before going to the main objective of this work, we intend to introduce one of the extensions of the teleparallel gravity on which the work is based.\\
 The different geometric setups distinguish teleparallel theory from General Relativity. The most peculiar attribute is using a flat affine connection (curvature free), known as Weitzenb$\ddot{o}$ck connection, instead of the usual Levi-Civita connection of the metric $g_{\mu \nu}$. The flatness of this connection gives freedom for a path-independent parallel transport, which is the reason behind the nomenclature "Teleparallel". The mathematical representation of Weitzenb$\ddot{o}$ck connection is,
 
 \begin{equation}
\label{connection}
\overset{w}{\Gamma}_{\nu \mu}^{\alpha} \equiv e_{\beta}^\alpha\partial_{\mu}e_{\nu}^\beta=-e_{\mu}^\beta \partial_{\nu} e_{\beta}^\alpha,
\end{equation}
where $e_{\beta}^\alpha$,  $e_{\nu}^\beta$ and $e_{\mu}^\beta$ are tetrads. The corresponding metric is defined as,
\begin{equation}
 \label{metric}
 g_{\mu \nu}=\eta_{\alpha \beta}\,\ e_{\mu}^{\alpha}\,\ e_{\nu}^{\beta},
 \end{equation}
where $\eta_{\alpha \beta}$ is the Minkowski metric. The above connection leads to the action for this theory in terms of the torsion scalar \eqref{torsionscalar}, which can be obtained using geometric tools like torsion tensor \eqref{torsion}, contorsion tensor \eqref{contorsion}, and superpotential tensor \eqref{superpotential}.

\begin{gather}
    T_{\mu \nu}^\alpha=\overset{w}{\Gamma}_{\nu \mu}^\alpha-\overset{w}{\Gamma}_{\mu \nu}^\alpha=e_{\beta}^\alpha(\partial_{\mu}e_{\nu}^\beta-\partial_{\nu}e_{\mu}^\beta),\label{torsion}\\
    {K^{\mu \nu}}_{\gamma} \equiv -\frac{1}{2}({T^{\mu \nu}}_{\gamma}-{T^{\nu \mu}}_{\gamma}-{T_{\gamma}}^{\mu \nu}),\label{contorsion}\\
    {S_{\gamma}}^{\mu \nu} \equiv \frac{1}{2}({K^{\mu \nu}}_{\gamma}+\delta_\gamma^\mu {T^{\alpha \nu}}_{\alpha}-\delta_\gamma^\nu {T^{\alpha \mu}}_{\alpha}),\label{superpotential}\\
    T\equiv  {S_{\gamma}}^{\mu \nu} T_{\mu \nu}^\gamma = \frac{1}{4}{T^{\gamma \mu \nu}T_{\gamma \mu \nu}}+ \frac{1}{2}{T^{\gamma \mu \nu}T_{\nu \mu \gamma}}-{T_{\gamma \mu}}^\gamma {T^{\nu \mu}}_\nu.\label{torsionscalar}
\end{gather}
It is well known that the source of modification for teleparallel action is GR action where Ricci scalar ($R$) is present. Using the Riemann curvature tensor and contorsion tensor, one can obtain the relation between the Ricci scalar and torsion scalar as $R=-T+B$. Here $B$ is the total divergence which acts as a boundary term in the TEGR action. The boundary term can be ignored as it does not contribute to the field equations. Finally, the TEGR action can be written as,
\begin{equation}
\label{TEGR}
    S_{TEGR}=-\frac{1}{16\pi G}{\int{{d^4}xe \,\ T}}.
\end{equation}

Similar to GR's $f(R)$ theory, TEGR can be generalized to the arbitrary function of $T$ as $f(T)$ theory. However, $f(T)$ and $f(R)$ will not remain equivalent due to the effect of nonlinearity in the boundary term. In this work, our interest lies in further extension of $f(T)$ which is the coupling between $T$ and trace of energy-momentum tensor ($\mathcal{T}$), termed as $f(T, \mathcal{T})$ theory \textbf{\cite{Ganiou:2015lvv, Pace:2017aon, Pace:2017dpu, Arouko:2020gbi, Moreira:2021uod}}. The corresponding action can be written as,
\begin{equation}
\label{action}
    S_{f(T, \mathcal{T})}=\frac{1}{16\pi G}{\int{{d^4}xe[T+f(T,\mathcal{T})]+\int{{d^4}xe\mathcal{L}_{m}}}},
\end{equation}
where $e=det(e_\mu^\alpha)=\sqrt{-g}$, $G$ is the Newton's constant and $\mathcal{L}_m$ is the matter Lagrangian. The field equations can be obtained by varying the action with respect to the tetrads,
\begin{multline}
 \label{field}
   (1+f_T)\left[e^{-1}\partial_\mu(e\,{e_\beta}^\alpha S_\alpha^{\gamma\mu})-e_\beta^\alpha T^{\mu}_{\nu \alpha} S_{\mu}^{\nu \gamma} \right]+
   \left(f_{TT}\, \partial_{\mu} T+ f_{T \mathcal{T}}\,\partial_{\mu} \mathcal{T}\right)e\,e_{\beta}^{\alpha} S_{\alpha}^{\gamma \mu}+e_{\beta}^\gamma\left(\frac{f+T}{4}\right)\\
-\frac{f_{\mathcal{T}}}{2} \left(e^{\alpha}_\beta \stackrel{em}{T}_\alpha^{\,\ \gamma} +p\, e_\beta^{\gamma} \right)= 4\pi G\, e^{\alpha}_\beta \stackrel{em}{T}_\alpha^{\,\ \gamma},
\end{multline}
with ${\stackrel{em}{T}_\alpha}^\gamma$ being the energy-momentum tensor, $f_T$, $f_\mathcal{T}$ being the first derivative with respect to $T$, $\mathcal{T}$, respectively and so on. For perfect fluid, ${\stackrel{em}{T}_\alpha}^\gamma$ has the form $diag (\rho_m,\, -p_m,\,-p_m,\,-p_m)$, $\rho_m$ and $p_m$ are the matter energy density and pressure, respectively. So the trace of energy-momentum tensor becomes $\mathcal{T}=\rho_m-p_m-p_m-p_m=\rho_m-3p_m$. The spatially flat Friedmann-Lemaitre-Robertson-Walker (FLRW) line element is to be assumed further. Its explicit representation is,
\begin{equation}
   \label{line}
       ds^2=dt^2-a^2(t)\delta_{ij}dx^i dx^j,
   \end{equation}
   where the scale factor $a(t)$ is in terms of time. For the above metric, $T=-6H^2$ and the tetrad field is considered as $e_\mu^\alpha=diag(1, a(t), a(t), a(t))$. Now, the modified Friedmann equations which will be used as the most important tools for further analysis, can be obtained using the above field equation and line element,
   \begin{equation}\
\label{friedmann1}
  {H^2=\frac{8\pi G}{3}{\rho_m}-\frac{1}{6}(f+12H^2f_T)+f_\mathcal{T}(\frac{\rho_m+p_m}{3})},
 \end{equation}
 
  \begin{equation}\
  \label{friedmann2}
  {\dot{H}=-4\pi G(\rho_m+p_m)}-\dot{H}(f_T-12H^2f_{TT})-H(\dot{\rho_m}-3{\dot{p_m}})f_{T\mathcal{T}}-f_\mathcal{T}(\frac{\rho_m+p_m}{2}).
  \end{equation}

  The motion equations can further be written in terms of dark energy components, considering $8 \pi G\equiv 1 $,
  \begin{gather}
    3H^2=\rho_m + \rho_{DE}, \label{fgr1}\\
    2\dot{H}=-(\rho_m + \rho_{DE}+ p_m + p_{DE}),\label{fgr2}
\end{gather}
where 
\begin{gather}
\rho_{DE}=-6 H^2 f_T -\frac{f}{2}+f_{\mathcal{T}}(\rho_m + p_m),\label{rd}
\end{gather}
\begin{equation}
p_{DE}= -\rho_{DE}+f_{\mathcal{T}} (\rho_m + p_m) -24 H^2 \dot{H} f_{TT}+2H(\dot{\rho_m}-3\dot{p_m})f_{T \mathcal{T}}+ 2\dot{H}f_T. \label{pd}
\end{equation}

Recent success of $f(T,\mathcal{T})$ theory as a fine alternative to the coherence model prompts us to review some prominent models of this theory. \cite{Harko:2014aja} considered the non-minimally coupled form $\alpha T^n \mathcal{T}+\Lambda $ and the quadratic form $\alpha \mathcal{T}+\gamma T^2$ to constrain through the energy conditions and found that the resulting models are consistent with the stability conditions. Two additive and multiplicative extensions of TEGR, $T+g(\mathcal{T})$ and $T\times g(\mathcal{T})$, have been considered by \cite{Saez-Gomez:2016wxb} to restrict by fitting the predicted distance modulus to that measured from type Ia supernovae. Further, the models are found to be good candidates to reproduce the late time acceleration. \cite{Mandal:2023bzo} attempted to alleviate the $H_0$ tension by performing statistical analysis on two new $f(T,\mathcal{T})$ models, $T e^{\,\alpha\, \frac{T_0}{T}}+\beta \mathcal{T}$ and $\alpha T_0(1-e^{-\beta \sqrt{T/T_0}}) + \gamma \mathcal{T}$. The deviation of their resulting $H_0$ value from $Planck2018$, $SH_0ES$, and $H_0LiCOW$ experiments are analyzed, leading to partially solving the tension. The interesting results of the theory develop inquisitiveness to look for the ultimate model that can describe evolution. In this regard, in the next section, we adopt a generic family of models without relying on any particular model to constrain them through the cosmographic approach.

\section{Cosmographical analysis}\label{sec:III}
We start this section with the basic assumptions of cosmographic parameters.
The Taylor series expansion of the scale factor $a(t)$ at the present time $t_0$ \textbf{\cite{Visser:2004bf,Visser:2015iua}} is
\begin{equation}\label{eq:a(t)}
    a(t)= 1+ \left.\sum_{n=1}^\infty \frac{1}{n!} \frac{d^na}{dt^n}\right|_{t=t_0} (t-t_0)^n.
\end{equation}
The coefficients of $(t-t_0)^n$ in the above series are the cosmographic parameters. We are going to mention the leading cosmographic parameters which will be used further \textbf{\cite{Poplawski:2006na, Poplawski:2006ew}}.
\begin{gather}
   \label{eq:H(t)}Hubble: H(t)=\frac{1}{a} \frac{da}{dt},\\
   \label{eq:q(t)}Deceleration: q(t)=-\frac{1}{aH^2}\frac{d^2a}{dt^2},\\
   \label{eq:j(t)}Jerk: j(t)=\frac{1}{aH^3}\frac{d^3a}{dt^3},\\
   \label{eq:s(t)}Snap: s(t)=\frac{1}{aH^4}\frac{d^4a}{dt^4}.
\end{gather}
The above parameters are entirely free from any cosmological models, which makes our task easy to execute this work in a model-independent approach. By straightforward calculation on the above equations, one can find the derivatives of Hubble parameters involving the cosmographic parameters as \textbf{\cite{Xu:2010hq, Luongo:2013rba}},
\begin{gather}
\begin{aligned}
    \dot{H}&=-H^2(1+q),\\
    \ddot{H}&=H^3(j+3q+2),\\
    \dddot{H}&=H^4(s-4j-3q(q+4)-6),
\end{aligned}
\end{gather}
 From now on dot$(\dot{})$ will always denote the derivative with respect to cosmic time $(t)$. Furthermore, one can obtain the derivatives of the torsion scalar and trace of energy-momentum tensor at present time $t=t_0$ using the above expressions as
 \begin{gather}
 \begin{aligned}
     T_0&=-6H_0^2,\\
     \dot{T}_0&=12 H_0^3 (q_0+1),\\
     \ddot{T}_0&=-12 H_0^4 (j_0+q_0(q_0+5)+3),\\
     \dddot{T}_0&=12 H_0^5 (j_0(3 q_0+7)+3 q_0 (4 q_0+9)-s_0+12),
 \end{aligned}
\end{gather}
and 
\begin{gather} 
\begin{aligned}
    \mathcal{T}_0&=3  H_0^2 \Omega_{m0},\\
     \dot{\mathcal{T}}_0&=-9 H_0^3  \Omega _{m0}, \\
     \ddot{\mathcal{T}}_0&=9  
 H_0^4 (q_0+4) \Omega _{m0}, \\
     \dddot{\mathcal{T}}_0&= -9 H_0^5 (j_0+12 q_0+20) \Omega _{m0}.
\end{aligned}
\end{gather}

The ultimate motivation behind finding all these terms and their derivatives is to find the Taylor series expansion of arbitrary function of the $f(T, \mathcal{T})$ theory. Now we define the Taylor series expansion around the point $(T_0,\mathcal{T}_0)$,
\begin{widetext}
\begin{multline}
    f(T,\mathcal{T})=f(T_0,\mathcal{T}_0)+\frac{1}{2!}f_T(T_0,\mathcal{T}_0)(T-T_0)+\frac{1}{2!}f_{\mathcal{T}}(T_0,\mathcal{T}_0)(\mathcal{T}-\mathcal{T}_0)+\frac{1}{3!}f_{TT}(T_0,\mathcal{T}_0)(T-T_0)^2+\\\frac{1}{3!}f_{T \mathcal{T}}(T_0,\mathcal{T}_0)(T-T_0)(\mathcal{T}-\mathcal{T}_0)+\frac{1}{3!}f_{\mathcal{T} \mathcal{T}}(T_0,\mathcal{T}_0)(\mathcal{T}-\mathcal{T}_0)^2+\mathcal{O},
\end{multline}
\end{widetext} 
where $\mathcal{O}$ is the term containing higher-order derivatives. For simplification, we assume a minimally coupling form of the theory as $f(T,\mathcal{T})=g(T)+h(\mathcal{T})$. So the $f_{T \mathcal{T}}$ terms will be reduced to zero in the above series. One can also consider some other kind of generic functional types, such as $g(T)\times h(\mathcal{T})$ or $exp(g(T)+ h(\mathcal{T}))$ or any such non-minimal form. Now in terms of the new function $g(T)$ and $h(\mathcal{T})$, we rewrite the series upto $4^{th}$ degree,
\begin{widetext}
\begin{multline}\label{eq:series}
    f(T,\mathcal{T})=g(T_0)+h(\mathcal{T}_0)+\frac{1}{2}g_T(T_0)(T-T_0)+\frac{1}{2}h_\mathcal{T}(\mathcal{T}_0) (\mathcal{T}-\mathcal{T}_0)+ \\ \frac{1}{6}g_{TT}(T_0)(T-T_0)^2+\frac{1}{6}h_{\mathcal{T} \mathcal{T}}(\mathcal{T}_0) (\mathcal{T}-\mathcal{T}_0)^2 + \frac{1}{24}g_{TTT}(T_0)(T-T_0)^3+\frac{1}{24}h_{\mathcal{T} \mathcal{T} \mathcal{T}}(\mathcal{T}_0) (\mathcal{T}-\mathcal{T}_0)^3+ \\ \frac{1}{120}g_{TTTT}(T_0)(T-T_0)^4+\frac{1}{120}h_{\mathcal{T} \mathcal{T} \mathcal{T} \mathcal{T}}(\mathcal{T}_0) (\mathcal{T}-\mathcal{T}_0)^4.
\end{multline}
\end{widetext} 
 Now, inserting the assumed form $g(T)+h(\mathcal{T})$ in the motion equations \eqref{friedmann1} and \eqref{friedmann2} at the present time gives us
\begin{equation}
\label{eq:motion3}
    6 H_0^2 \left(2 g^{(1)}+1\right)+g+h-2 \mathcal{T}_0 \left(h^{(1)}+1\right)=0,
\end{equation}
\begin{equation}
\label{eq:motion4}
    -A H_0^2(1+q_0) + \mathcal{T}_0 \left(h^{(1)}+1\right)=0
\end{equation}
with $A=2 \left(g^{(1)}-12 H_0^2 g^{(2)}+1\right)$, $g\equiv g(T_0)$, $h\equiv h(\mathcal{T}_0)$, $g^{(1)}\equiv g_T(T_0)$, $g^{(2)}\equiv g_{TT}(T_0)$ and $h^{(1)}\equiv h_\mathcal{T}(\mathcal{T}_0)$. The derivatives will be denoted in this manner in further work. The following equations (\ref{x}, \ref{y}) are obtained by taking further derivatives of the second motion equation \eqref{eq:motion4}.
\begin{widetext}
\begin{multline}
\label{x}
   H_0 \left(2 g^{(1)} (j_0+3 q_0+2)+ 3 H_0^2 \left(-8 g^{(2)} (j_0+3 q_0(q_0+3)+5)+96 H_0^2 (q_0+1)^2 g^{(3)}-  9 h^{(2)} \Omega _{m0}^2\right) \right. \\ \left. -9 (h^{(1)}+1) \Omega _m{0}+2 (j_0+3 q_0+2)\right)=0,
\end{multline}

\begin{multline}\label{y}
     288 H_0^4 (q_0+1) \left(g^{(3)} (j_0+3 q_0 (2 q_0+5)+8)-12 H_0^2 (q_0+1)^2 g^{(4)}\right)=\\
     2 (g^{(1)}+1) (-4 j_0-3 q_0 (q_0+4)+s_0-6)+48 H_0^2 (q_0+1) (j_0+3 q_0+2) \left(g^{(2)}-12 H_0^2 g^{(3)}\right)\\+24 H_0^2 g^{(2)} (j_0 (7 q_0+11)+q_0 (3 q_0 (q_0+11)+56)-s_0+23)+ 9 (q_0+4) \Omega _{m0} \\\times \left(3 H_0^2 h^{(2)} \Omega _{m0}+h^{(1)}+1\right)+81 H_0^2 \Omega _{m0}^2 \left(3 H_0^2 h^{(3)} \Omega _{m0}+2 h^{(2)}\right).
\end{multline}

By solving the equations [\ref{eq:motion4}-\ref{y}], one can find $q_0$, $j_0$ and $s_0$ in terms of the unknown parameters,

\begin{equation}\label{eq:q0}
    q_0=\frac{1}{A}\left(3 (h^{(1)}+1) \Omega _{m0}\right)-1,
\end{equation}

\begin{multline}\label{eq:j0}
    j_0=\frac{1}{A}\left(-4 - 6 q_0 - 2 g^{(1)} (2 + 3 q_0) + 
   9 (1 + h^{(1)}) \Omega _{m0} \right.\\\left.
   +3 H_0^2 (-96 g^{(3)} H_0^2 (1 + q_0)^2 + 
      8 g^{(2)} (5 + 3 q_0 (3 + q_0)) + 9 h^{(2)} 
{\Omega _{m0}}^2\right),
\end{multline}

\begin{align}
    \begin{split}\label{eq:s0}
    s_0=&-\frac{1}{A}\left(-2 g^{(1)} (4 j_0+3 q_0 (q_0+4)+6)+4 \left(162 H_0^2 g^{(2)}-864 H_0^4 g^{(3)}+864 H_0^6 g^{(4)}+9 (h^{(1)}+1) \Omega _{m0}-2 j_0-6 q_0-3\right) \right. \\&\left. -6 \left(-4 H_0^2 g^{(2)} (j_0 (9 q_0+13)+3 q_0 (q_0+2) (q_0+11))+144 H_0^4 g^{(3)} \left((j_0+11) q_0+j_0+2 q_0^3+9 q_0^2\right) \right.\right.\\&\left. \left. +q_0 \left(q_0-576 H_0^6 (q_0 (q_0+3)+3) g^{(4)}\right)\right)+27 H_0^2 (q_0+10) h^{(2)} \Omega _{m0}^2+243 H_0^4 h^{(3)} \Omega _{m0}^3+9 q_0 (h^{(1)}+1) \Omega _{m0} \right).
\end{split}
\end{align}

\end{widetext}

\begin{figure*}
    \centering
    \includegraphics[width=0.8\linewidth]{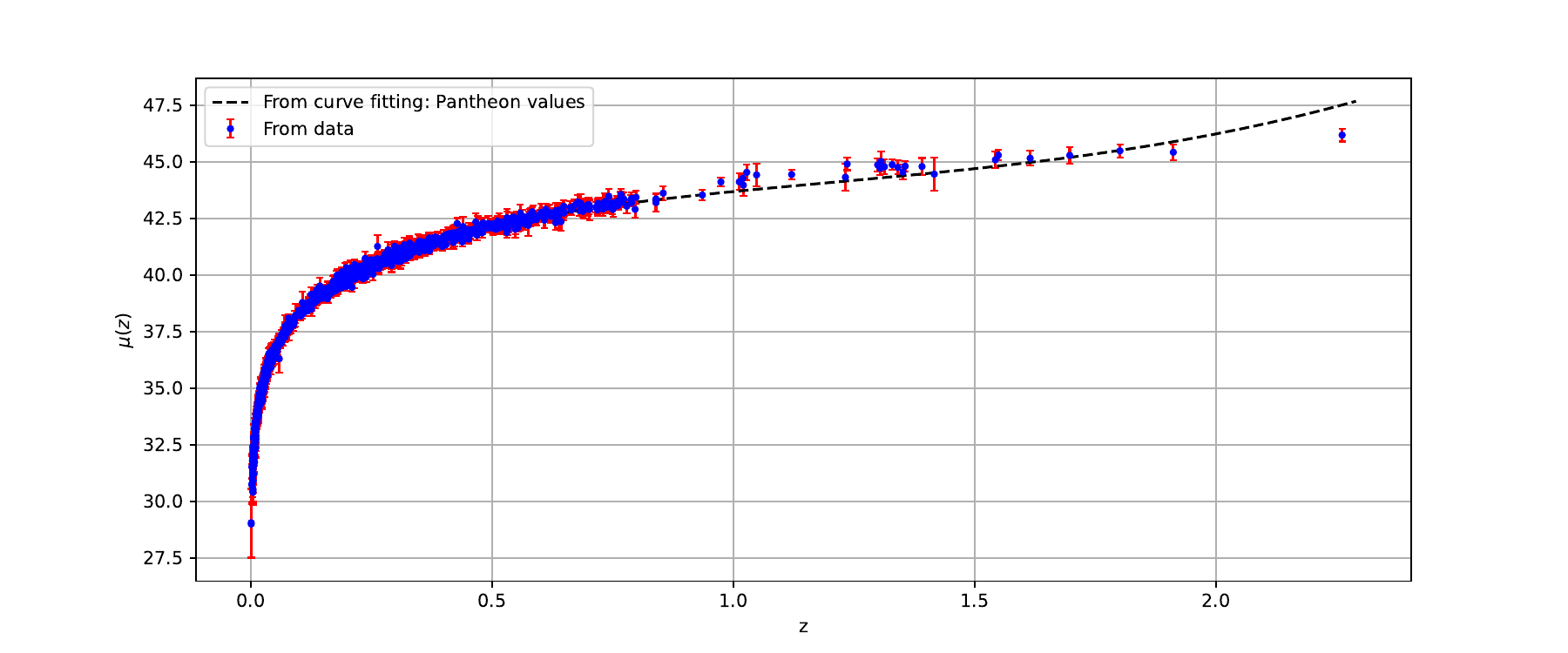}
    \caption{Distance modulus function: Plot illustrating the $\mu(z)$ profile resulting from the MCMC analysis of parameters $(H_0,g^{(1)},g^{(2)},g^{(3)},g^{(4)},h^{(1)},h^{(2)},h^{(3)},\Omega_{m0})$ using  Pantheon+SH0ES dataset. The error bars represent 1701 data points of the Pantheon+SH0ES sample.}
    \label{fig:muz}
\end{figure*}

\begin{figure*}
    \centering
    \includegraphics[width=\linewidth]{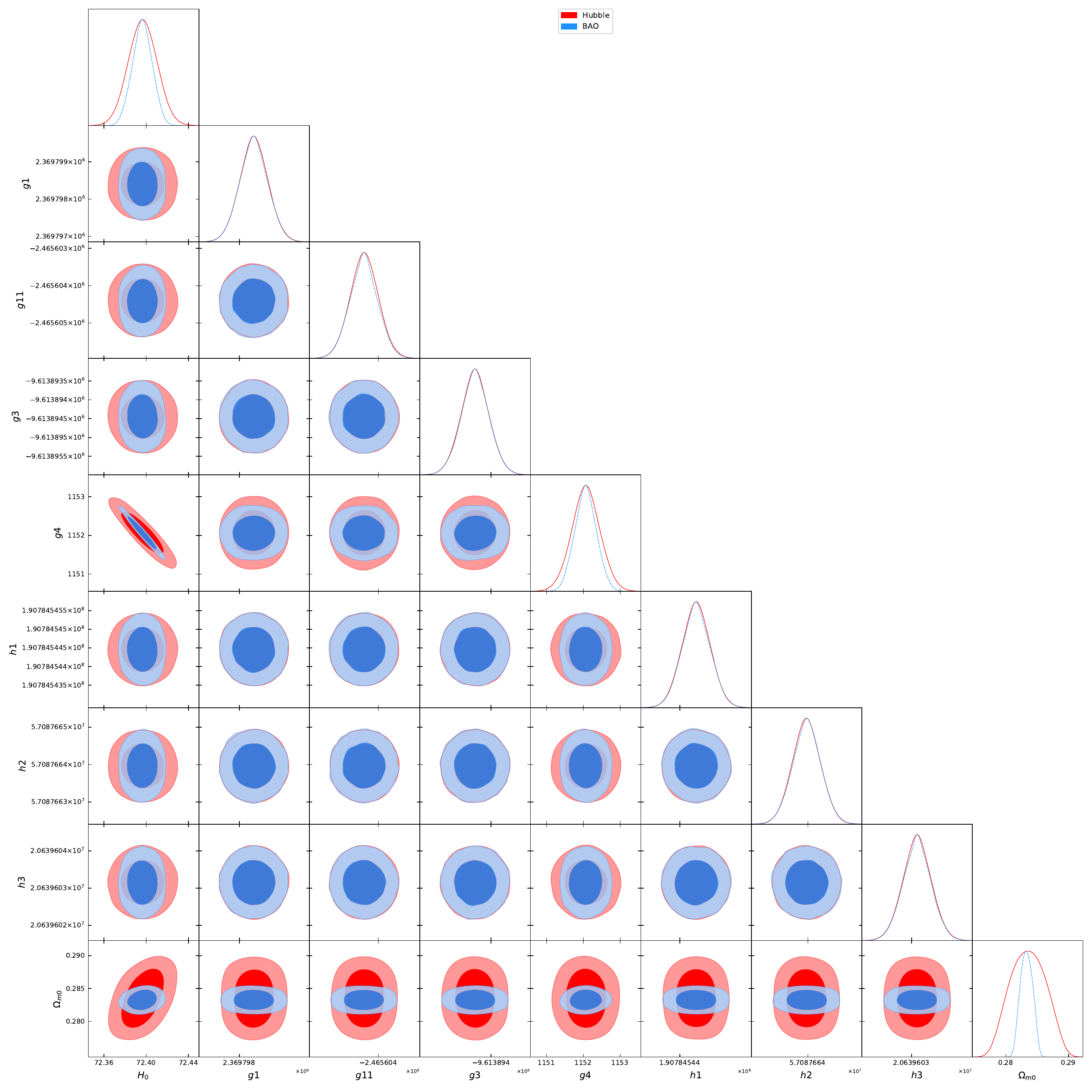}
    \caption{2D likelihood contours obtained from MCMC analysis of CC and BAO datasets}
    \label{fig:hubble_bao}
\end{figure*}

\begin{figure*}
    \centering
    \includegraphics[width=\linewidth]{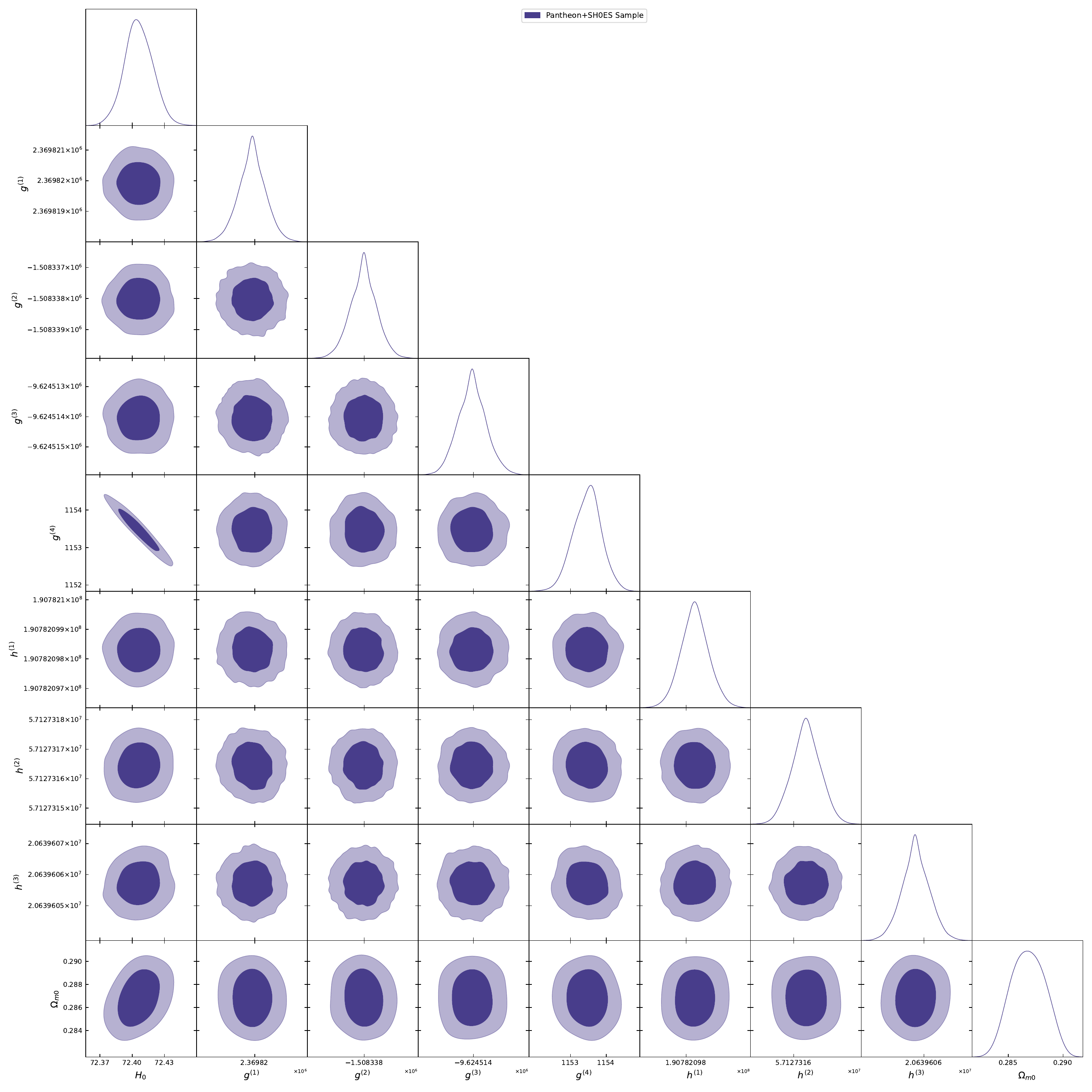}
    \caption{2D likelihood contours obtained from MCMC analysis of Pantheon+SH0ES dataset.}
    \label{fig:pantheon}
\end{figure*}

\section{Observational constraints}\label{sec:IV}

\subsection{Cosmographic description of parameters}

Thus far, we have examined the present-time assessment of cosmographic parameters. These evaluations serve as the basis for our analysis in understanding various cosmological parameters in terms of redshift. To begin, let us first revisit the relationship between redshift and emission time. We start with the standard relation $a(t)=\frac{1}{1+z}$, with $a(t_0)=1$. By employing the Taylor series expansion of $a(t)$ in terms of cosmographic parameters (see equations \ref{eq:a(t)} - \ref{eq:s(t)} ), one can obtain the series expansion of $t(z)$. This methodology proves instrumental in deducing expansions for other parameters in terms of redshift.

The Hubble parameter is a key concept in contemporary cosmology and an essential aspect of our understanding of the dynamics and evolution of the Universe. It represents the rate of expansion of the Universe, illustrating the relationship between the velocity at which distant galaxies are receding from us and their respective distances. Furthermore, it plays a crucial role in the development of the cosmic distance ladder, which is a hierarchical set of techniques used for calculating distances to astronomical objects through cosmological observations. Using the cosmographic technique to derive the series expansion of emission time and employing the definition of $H$ \eqref{eq:H(t)}, along with \eqref{eq:s(t)}, yields \textbf{\cite{Gao:2024}}

\begin{equation}\label{eq:H(z)}
    H(z) = H_0 \left[1+\alpha_1 z+\alpha_2 z^2+\alpha_3 z^3+\xi_H\right],
\end{equation}
where $\xi_H$ represents the higher order terms of $H(z)$ and $\alpha_i$'s are coefficients in terms of cosmographic parameter given by 

\begin{align*}
    \alpha_1&=(1+q_0),\\
    \alpha_2&=\frac{1}{2!}(j_0-q_0^2),\\
    \alpha_3&=-\frac{1}{3!}(s_0+3j_0+4j_0q_0-3q_0^2-3q_0^3).
\end{align*}

An important quantity that quantifies the relationship between the angular size of an object and its physical size when observed from Earth is the angular diameter distance $d_A$. It enables us to determine the actual size of distant objects from their apparent angular size and is essential for measuring the separations between galaxies, galaxy clusters, and other cosmic formations. This is represented by the relation \textbf{\cite{Tonghua:2023hdz}}

\begin{equation}\label{eq:d_A}
    d_A(z) = \frac{d_L(z)}{(1+z)^2}.
\end{equation}

It shows how the apparent brightness of distant celestial objects is affected by the expansion of the Universe. Here, $d_L(z)$ represents the luminosity distance expressed in megaparsecs (Mpc), which accounts for the redshift caused by the Universe's expansion and denotes the distance at which an object with a known intrinsic luminosity would appear. Mathematically, it is described as \textbf{\cite{Busti:2015xqa, Demianski:2012ra, Piedipalumbo:2015jya}}

\begin{equation}\label{eq:d_L}
    d_L(z) = \frac{c(1+z)}{H_0}\int_0^z \frac{d\xi}{E(\xi)}.
\end{equation}

In this equation, $c$ signifies the speed of light, $H_0$ is the Hubble constant, and $E(z)$ is defined as the ratio of the Hubble parameter at redshift $z$ to the present-day Hubble constant. This can calibrated using equation \eqref{eq:H(z)}. Thus, the series expansion of \eqref{eq:d_L} is given by

\begin{equation}\label{eq:d_L(z)}
    d_L(z) = \frac{cz}{H_0} \left[1+\beta_1 z+\beta_2 z^2+\beta_3 z^3+\xi_{d_L}\right],
\end{equation}

where $\xi_{d_L}$ denotes the higher-order terms of $d_L(z)$, while the coefficients $\beta_i$ are expressed in terms of cosmographic parameters, as follows

\begin{align*}
    \beta_1&=\frac{1}{2!}(1-q_0),\\
    \beta_2&=-\frac{1}{3!}(1-q_0-3q_0^2+j_0),\\
    \beta_3&=\frac{1}{4!}(2-2q_0-15q_0^2-15q_0^3+5j_0+10q_0j_0+s_0).\\
\end{align*}

In light of \eqref{eq:d_L(z)} and the Taylor series expansion of $(1+z)^{-2}$, \eqref{eq:d_A} can be reformulated as

\begin{equation}
    d_A(z) = \frac{cz}{H_0} \left[1+\gamma_1 z+\gamma_2 z^2+\gamma_3 z^3+\xi_{d_A}\right].
\end{equation}

Here, $\xi_{d_A}$ represents the higher-order terms of $d_A(z)$, with the coefficients $\gamma_i$ specified in terms of cosmographic parameters, as outlined below

\begin{align*}
    \gamma_1&=-\frac{1}{2!}(3+q_0),\\
    \gamma_2&=\frac{1}{3!}(11+7q_0+3q_0-j_0),\\
    \gamma_3&=-\frac{1}{4!}(50+46q_0+39q_0^2+15q_0^3-13j_0-10q_0j_0-s_0).\\
\end{align*}

Equation \eqref{eq:d_L} enables us to compute the distance modulus $\mu$, which represents the difference in magnitude between the apparent magnitude $(m)$ and the absolute magnitude $(M)$. The distance modulus proves useful in estimating the distances of objects based on observed brightness when direct distance measurements are not feasible. The theoretical expression for distance modulus is given by

\begin{equation}\label{eq:mu}
    \mu(z) = 5\log_{10}\left(\frac{d_L(z)}{1\ \text{Mpc}} \right) + 25.
\end{equation}

Thus, series expansion of the above function can be provided as

\begin{equation}\label{eq:mu(z)}
    \mu(z)=\frac{5}{\log 10} \left[\log z+\delta_1 z+\delta_2 z^2+\delta_3 z^3+\xi_{\mu}\right].
\end{equation}

In this expression, $\xi_\mu$ denotes the higher-order terms of $\mu(z)$, while the coefficients $\delta_i$ represent parameters within the cosmographic framework, defined as

\begin{align*}
    \delta_1&=-\frac{1}{2}(-1+q_0),\\
    \delta_2&=-\frac{1}{24}(7-10q_0-9q_0^2+4j_0),\\
    \delta_3&=\frac{1}{24}(5-9q_0-16q_0^2-10q_0^3+7j_0+8q_0j_0+s_0).
\end{align*}

\begin{table*}
 \centering
 \caption{ Best fit range of the parameters with $1-\sigma$ confidence level for constrained parameters.
 }
 
 \label{table1}
    \begin{tabular}{|c||c|c|c|}
    \hline
    
          & $CC$ & $BAO$ & $Pantheon+SH0ES$ \\
    \hline
    \hline

    $H_0$ & $(72.424,72.450)$ & $(72.387,72.405)$ & $(72.392,72.418)$ \\
    \hline
    
     $g^{(1)}$ & $(2371730.05,2371730.85)$ & $(2369798.02,2369798.82)$ & $(2369819.41,2369820.39)$  \\
    \hline
    
     $g^{(2)}$ & $(-2250802.13,-2250801.35)$ & $(-2465604.79,-2465603.99)$ & $(-1508338.51,-1508337.51)$  \\
    \hline
    
     $g^{(3)}$ & $(-9457280.63,-9457279.83)$ & $(-9613894.84,-9613894.04)$ &  $(-9624514.53,-9624513.53)$\\
    \hline
    
     $g^{(4)}$ & $(1131.8,1132.56)$ & $(1151.77,1152.35)$ & $(1153.09,1153.89)$ \\
    \hline
    
     $h^{(1)}$ & $(191000553,191000553.80)$ & $(190784544.07,190784544.87)$ & $(190782097.84,190782098.82)$ \\
    \hline
    
     $h^{(2)}$ & $(56396393.89,56396394.69)$ & $(57087663.57,57087664.37)$ & $(57127315.97,57127316.93)$ \\
    \hline
    
     $h^{(3)}$ & $(20639685.58,20639686.38)$ & $(20639602.75,20639603.55)$ & $(20639605.24,20639606.20)$ \\
    \hline
    
     $\Omega_{m0}$ & $(0.2815,0.2867)$ & $(0.28225,0.28421)$ & $(0.2853,0.2883)$ \\
    \hline
   
    \end{tabular}
\end{table*}

\subsection{Data Sets and Methodology}

To reconstruct $f(T,\mathcal{T})$ gravity and accurately reflect the dynamics of our Universe, it is essential to make use of observational data and employ suitable methodologies for parameter estimation. In this section, we provide an overview of the observational datasets utilized and the methodology adopted to constrain the model parameters. 

In our study, we have considered the general minimal form of $f(T,\mathcal{T})$ coupling in the form of $g$ and $h$, where the latter is a pure function of $\mathcal{T}$ and the former is a pure function of $T$ alone. The Taylor series expansion of the same is provided in \eqref{eq:series}. Since we are relying on a model-independent approach, $g$ and $h$ are not assigned specific functional forms. Thus, the present values of these functions and their derivatives are unknown. To this end, we consider $g^{(1)}, g^{(2)}, g^{(3)}, g^{(4)}, h^{(1)}, h^{(2)}$, and $h^{(3)}$ as free parameters whose values need to be calibrated for the present epoch. These parameters along with $H_0$ and $\Omega_{m0}$ are constrained using the statistical Bayesian approach.

To determine the best-fit values for the parameters, we minimize the $\chi^2$ function. Given the relationship between $\chi^2$ and likelihood as $\mathcal{L} \propto e^{-\frac{\chi^2}{2}}$, minimizing $\chi^2$ is equivalent to maximizing the likelihood, which is further equivalent to minimizing the negative log-likelihood. To obtain constraints on the free parameters, we utilize the Markov Chain Monte Carlo (MCMC) sampling method. In order to achieve this we employ the following data sets:

\subsubsection{Cosmic Chronometers data} The Cosmic Chronometers (CC) method provides measurements of the Hubble rate by examining the oldest and least active galaxies, which are closely spaced in terms of redshift, utilizing the differential aging technique. This approach relies on the definition of the Hubble rate $ H=-\frac{1}{1+z}\frac{dz}{dt}$ for an FLRW metric. The significance of the CC method lies in its capacity to determine the Hubble parameter without relying on any specific cosmological assumption. This makes it a vital aid for analyzing cosmological models. Here, we have made use of 31 CC data points within the redshift range $(0.1,2)$ obtained from various surveys \cite{Jimenez:2003iv,Simon:2004tf,Stern:2009ep,Moresco:2012jh,Zhang:2012mp,Moresco:2015cya,Moresco:2016mzx,Ratsimbazafy:2017vga}. For CC data the $\chi^2$ function can be written as
\begin{equation}
    \chi_{CC}^2 = \Delta H^T (C_{CC}^{-1})\Delta H,
\end{equation}
where $\Delta H$ is the vector containing the difference between theoretical and observed values of $H(z)$ for each redshift data and $C_{CC}$ is the covariance matrix elements being errors associated with the observed $H$ values. The theoretical value of the Hubble parameter is computed using the equation \eqref{eq:H(z)}.

\subsubsection{Baryonic Acoustic Oscillations data}
Baryonic Acoustic Oscillations (BAO) serve as a prominent observational tool for examining the large-scale structure of the Universe. These oscillations originate from acoustic waves in the early Universe, which cause compression in the baryonic matter and radiation within the photon-baryon fluid. This compression leads to a distinctive peak in the correlation function thus acting as a standard scale for measurement of distances across the Universe. Here we employ 26 independent line-of-sight measurements of BAO data points \cite{BOSS:2014hwf,Gaztanaga:2008xz,Blake:2012pj,Chuang:2013hya,Chuang:2012qt,Busca:2012bu,Oka:2013cba,BOSS:2013rlg,BOSS:2013igd,Bautista:2017zgn,BOSS:2016zkm,Ross:2014qpa} for which the $\chi^2$ function is given by

\begin{equation}
    \chi^2_{BAO} = \sum_{i=1}^{26}\left[\frac{(H_{th}(z_i,\Theta)-H_{ob}(z_i))^2)}{\sigma_H^2(z_i)}\right].
\end{equation}
Similar to the CC method, the theoretical value of the Hubble parameter is computed using its cosmographic expansion.

\begin{figure}[b]
    \centering
    \includegraphics[width=0.6\linewidth]{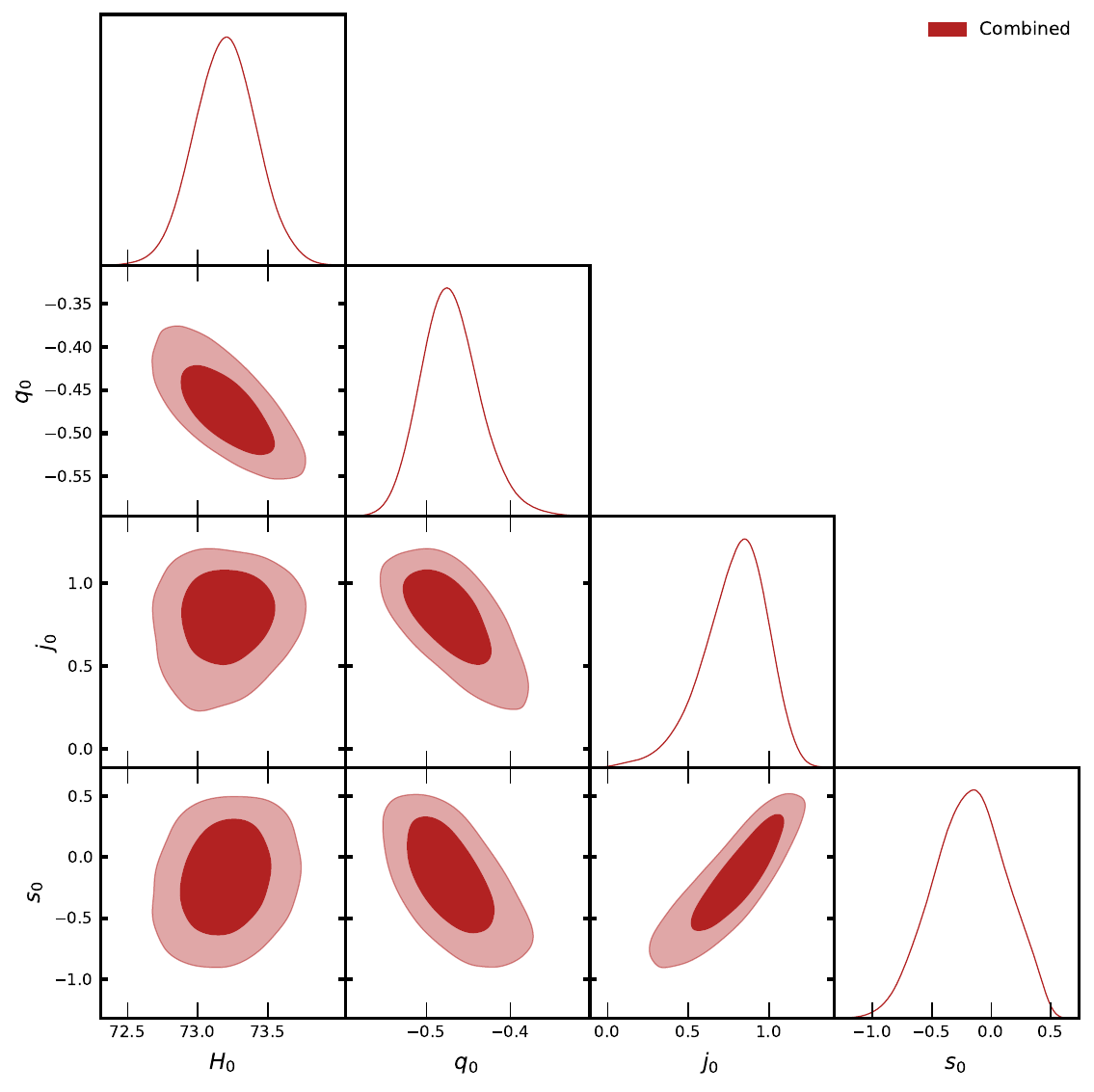}
    \caption{2D likelihood contours obtained from MCMC analysis of combined CC, BAO, Pantheon+SH0ES dataset.}
    \label{fig:qjs}
\end{figure}

\subsubsection{Pantheon+SH0ES data} 
The Pantheon+SH0ES Data dataset, (See reference \cite{Riess:2021jrx,Malekjani:2023dky,Brout:2022vxf,Brout:2021mpj,Scolnic:2021amr}), encompasses distance moduli calculated from 1701 light curves of 1550 Type Ia supernovae (SNeIa) gathered from 18 different surveys. These light curves span a redshift range of $0.001\leq z\leq 2.2613$. Notably, within the dataset, 77 light curves correspond to galaxies hosting Cepheids. A primary advantage of Pantheon+SH0ES Data lies in its utility for constraining the Hubble constant $H_0$ in conjunction with other free parameters. We estimate the theoretical distance modulus for the SNeIa sample as defined in the equation \eqref{eq:mu(z)}. The free parameters $(H_0,g^{(1)},g^{(2)},g^{(3)},g^{(4)},h^{(1)},h^{(2)},h^{(3)},\Omega_{m0})$ are constrained by utilizing equations \eqref{eq:q0}-\eqref{eq:s0} in \eqref{eq:mu(z)}. Consequently, the distance residual $\Delta\mu$ is written as

    \begin{equation}
        \Delta\mu_i=\mu_i-\mu_{th}(z_i),
    \end{equation}
    
    When examining data from the SNeIa sample, a degeneracy arises between the parameters $H_0$ and $M$. To overcome this, a modification is introduced to the SNeIa distance residuals \cite{Perivolaropoulos:2023iqj,Brout:2022vxf} as
    
    \begin{equation}
        \Delta\Tilde{\mu}=\begin{cases}
    			\mu_i-\mu_i^{Ceph}, & \text{if $i$ belongs to Cepheid hosts}\\
                \mu_i-\mu_{\text{th}}(z_i), & \text{otherwise}
    		 \end{cases}
    \end{equation}
    
    Here, $\Delta\Tilde{\mu}$ is the modified distance residual, and $\mu_i^{Ceph}$ represents the distance modulus of the Cepheid host of the $i^{th}$ SNeIa. The $\chi^2$ function for SNeIa is provided as

    \begin{equation}\label{Eq:ChiSN}
        \chi^2_{SNeIa}= \Delta\mu^T (C_{\text{stat+sys}}^{-1})\Delta\mu.
    \end{equation}

\begin{table}
\caption{Best fit range of the parameters with $1-\sigma$ confidence level for the combined dataset.}
    \centering
    \begin{tabular}{|c||c|}
    \hline
        Parameter & Range \\
        \hline\hline
         $H_0$ & $(72.98,73.42)$ \\\hline
         $q_0$ & $(-0.509,-0.44)$ \\\hline
         $j_0$ & $(0.63,1.01)$ \\\hline
         $s_0$ & $(0.48,0.12)$ \\\hline
    \end{tabular}
    
    \label{tab:combined}
\end{table}

\begin{figure}[]
    \centering
    \includegraphics[width=0.4\linewidth]{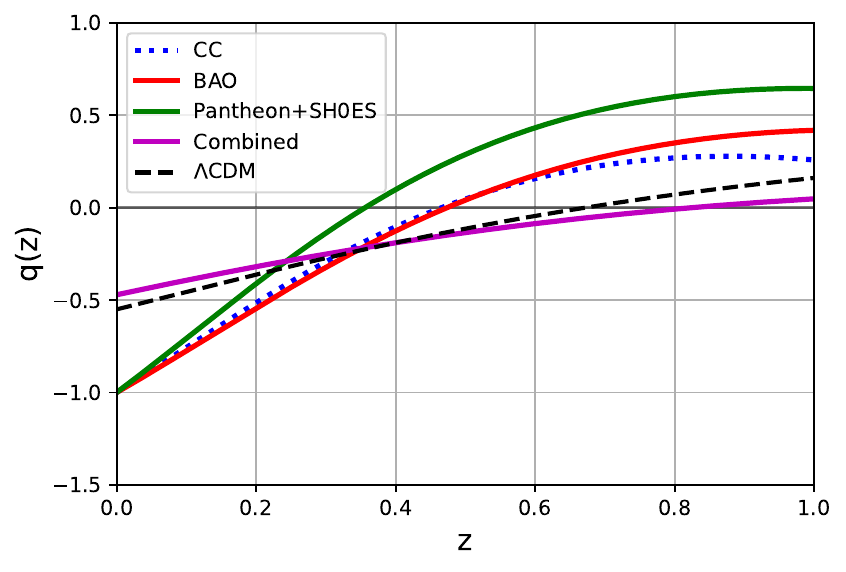}
    \caption{Deceleration Parameter: Plot illustrating the $q(z)$ profile resulting from the MCMC analysis of parameters $(H_0,g^{(1)},g^{(2)},g^{(3)},g^{(4)},h^{(1)},h^{(2)},h^{(3)},\Omega_{m0})$ using (\mycirc[blue]) CC, (\mycirc[red]) BAO, and (\mycirc[ForestGreen]) Pantheon+SH0ES datasets, and using parameters $(H_0, q_0, j_0, s_0)$ for (\mycirc[magenta]) the combined dataset. For (\mycirc) $\Lambda$CDM, the curve is plotted with $\Omega_{m_0}=0.3$, $\Omega_{\Lambda_0}=0.7$ and $H_0=67.8~\text{km}~\text{s}^{-1}~\text{Mpc}^{-1}$}
    \label{fig:q}
\end{figure}

\subsection{Results}
This section is dedicated to discuss the outcomes obtained from curve fitting and MCMC using various datasets.  Further, we verify the credibility of those results by plotting cosmological parameters. From \autoref{fig:muz}, we depict that the distance modulus function is in excellent agreement with 1701 data points of Pantheon+SH0ES sample. The 2D contours for all the datasets up to $2-\sigma$ CL are presented in plots \ref{fig:hubble_bao} and \ref{fig:pantheon}.  We summarize the best-fit $1-\sigma$ range in \autoref{table1}. It is worth to note that a considerable deviation in ranges for $1-\sigma$ CL can be observed from MCMC analysis of all three datasets, while for $2-\sigma$ CL, parameters lie in almost similar ranges. In addition, we conducted MCMC analysis with $H_0$, $q_0$, $j_0$, and $s_0$ as free variables. To accomplish this, we utilized the combined dataset (CC+BAO+Pantheon+SH0ES), and the corresponding 2D likelihood contour up to $2-\sigma$ is depicted in \autoref{fig:qjs}. These values are summarized in \autoref{tab:combined}.  

\begin{figure}[]
    \centering
    \includegraphics[width=0.4\linewidth]{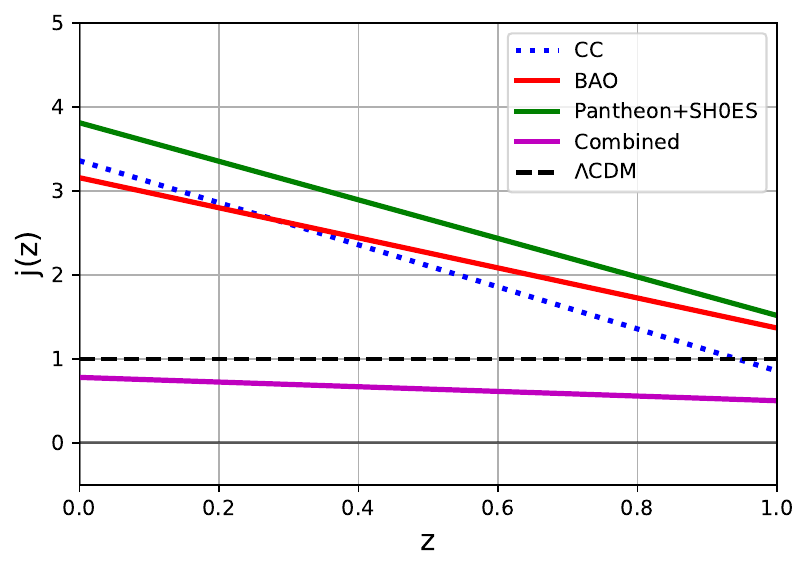}
    \caption{Jerk Parameter:  Plot illustrating the $j(z)$ profile resulting from the MCMC analysis of parameters $(H_0,g^{(1)},g^{(2)},g^{(3)},g^{(4)},h^{(1)},h^{(2)},h^{(3)},\Omega_{m0})$ using (\mycirc[blue]) CC, (\mycirc[red]) BAO, and (\mycirc[ForestGreen]) Pantheon+SH0ES datasets, and using parameters $(H_0, q_0, j_0, s_0)$ for (\mycirc[magenta]) the combined dataset. For (\mycirc) $\Lambda$CDM, the curve is plotted with $\Omega_{m_0}=0.3$, $\Omega_{\Lambda_0}=0.7$ and $H_0=67.8~\text{km}~\text{s}^{-1}~\text{Mpc}^{-1}$}
    \label{fig:j}
\end{figure}

For combined analysis, the obtained ranges of $q_0$, $j_0$, and $s_0$ are comparable to $\Lambda$CDM. Through the deceleration parameter in \autoref{fig:q}, one can observe the dynamic phase transition from deceleration to acceleration at certain redshifts which are explicitly mentioned in \autoref{table2}. To describe the current accelerated expansion $q_0$ has to be in $-1 < q_0 \leq 0$, while $q_0=-1$ explains the de Sitter Universe, completely dominated by dark energy. The transition redshift ($z_t$) for $\Lambda$CDM is around $z_t \sim 0.7$. The redshifts we obtained from \autoref{fig:q} are well within the best-fit $1-\sigma$ ranges found from MCMC analysis in \cite{Muthukrishna:2016evq}. The jerk parameter obtained from the second-order derivative of the scale factor series should be positive at the present time.
  
\begin{figure}[]
    \centering
    \includegraphics[width=0.4\linewidth]{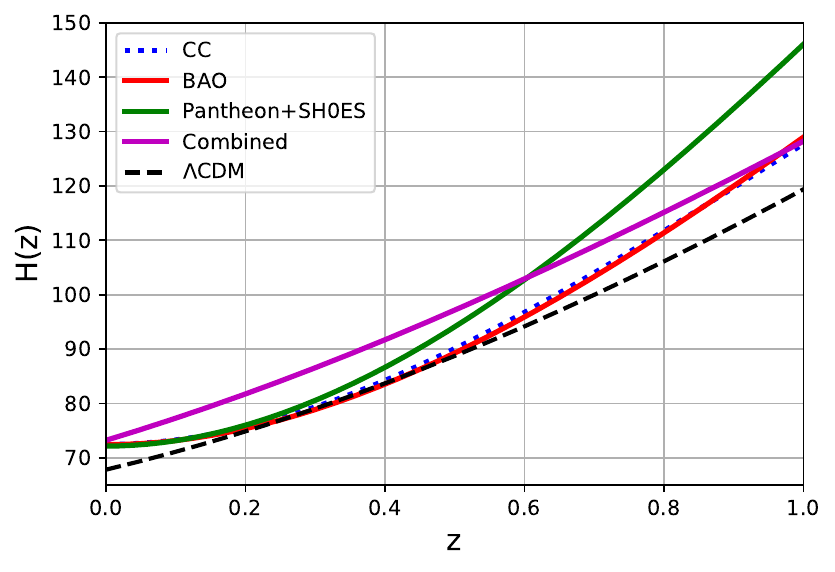}
    \caption{Hubble Parameter:  Plot illustrating the $H(z)$ profile resulting from the MCMC analysis of parameters $(H_0,g^{(1)},g^{(2)},g^{(3)},g^{(4)},h^{(1)},h^{(2)},h^{(3)},\Omega_{m0})$ using (\mycirc[blue]) CC, (\mycirc[red]) BAO, and (\mycirc[ForestGreen]) Pantheon+SH0ES datasets, and using parameters $(H_0, q_0, j_0, s_0)$ for (\mycirc[magenta]) the combined dataset. For (\mycirc) $\Lambda$CDM, the curve is plotted with $\Omega_{m_0}=0.3$, $\Omega_{\Lambda_0}=0.7$ and $H_0=67.8~\text{km}~\text{s}^{-1}~\text{Mpc}^{-1}$}
    \label{fig:H}
\end{figure}

\begin{figure}[]
    \centering
    \includegraphics[width=0.4\linewidth]{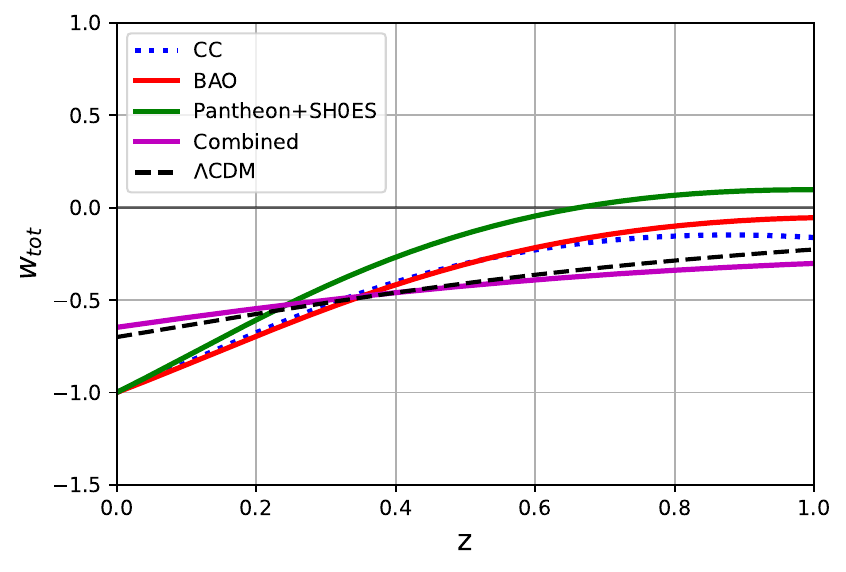}
    \caption{Effective EoS Parameter:  Plot illustrating the $w_{\text{eff}}$ profile resulting from the MCMC analysis of parameters $(H_0,g^{(1)},g^{(2)},g^{(3)},g^{(4)},h^{(1)},h^{(2)},h^{(3)},\Omega_{m0})$ using (\mycirc[blue]) CC, (\mycirc[red]) BAO, and (\mycirc[ForestGreen]) Pantheon+SH0ES datasets, and using parameters $(H_0, q_0, j_0, s_0)$ for (\mycirc[magenta]) the combined dataset. For (\mycirc) $\Lambda$CDM, the curve is plotted with $\Omega_{m_0}=0.3$, $\Omega_{\Lambda_0}=0.7$ and $H_0=67.8~\text{km}~\text{s}^{-1}~\text{Mpc}^{-1}$}
    \label{fig:w}
\end{figure}

  Though the behavior of the jerk parameter (see \autoref{fig:j}) obtained from the result of MCMC is slightly deviating from $\Lambda$CDM for which it is always $1$, positivity at the present redshift makes it a viable candidate to explain the acceleration. From \autoref{fig:H}, we found that for all individual samples along with the combined one, the Hubble parameter shows very similar behavior to the standard cosmological model. Furthermore, the total equation of state parameter (denoted as $w_{tot}$) is plotted against redshift in \autoref{fig:w}. From \autoref{table2} it can be seen that the present values of $w_{tot}$ are not subceeding $-1$, which confirms the quintessence behavior of the Universe.

\begin{table}
 \centering
 \caption{ Present values of the cosmographic parameters for the three datasets.
 }
 
 \label{table2}
 
    \begin{tabular}{|c||c|c|c|}
    \hline
    
          & $CC$ & $BAO$ & $Pantheon+SH0ES$ \\
    \hline
    \hline

    $H_0$ & 72.4361776
 & 72.3961722
 & 72.138
 \\
    \hline
    
     $q_0$ & -0.999426 & -0.999477 &  -0.999371 \\
    \hline
    
     $j_0$ & 3.359057 & 3.157094 & 3.811645  \\
    \hline
    
     $s_0$ & 5.855289 & 4.941443 & 6.098992 \\
    \hline
    
     $w_0$ & -0.999617 & -0.999651 & -0.999581 \\
    \hline
    
     $z_t$ & 0.465654 & 0.473282 & 0.355881 \\
    \hline

    \end{tabular}
\end{table}

\section{Interpretation and Conclusion}\label{sec:V}

 To reconstruct the theory, we started with the definition of the Taylor series function of $f(T,\mathcal{T})$. We have constrained the independent free parameters involved in the series, using that one can find the remaining ones such as $g(T_0)$, $h(\mathcal{T}_0)$ and $h_{\mathcal{T}\mathcal{T}\mathcal{T}\mathcal{T}}(\mathcal{T}_0)$. One can directly obtain $g(T_0)+h(\mathcal{T}_0)$ from the motion equation \eqref{eq:motion3}, by substituting the resulting values from MCMC. Also, the fourth derivative of $h$ can be computed by differentiating \eqref{eq:motion3} three times.

Cosmography is used as an essential tool to reconstruct a theory that already has a plethora of successful models describing various phases of the evolution of the Universe. All Taylor expansions related to cosmography are valid within the observable domain where $z \ll 1$, allowing one to establish constraints on the present-day universe. The $\mu(z)$ and $H(z)$ functions involving cosmographic parameters are used to constrain the $f(T,\mathcal{T})$ theory by using the Markov Chain Monte Carlo method for the SNeIa, BAO, and Hubble datasets. Also, we obtained an independent range for the parameters $H_0$, $q_0$, $j_0$, and $s_0$ using the combined dataset (CC+BAO+SNeIa). All ranges that the MCMC analysis yielded are summarized in tabular form in \autoref{table1}. Further, the constrained functions from all datasets are examined and compared through various cosmological parameters against low redshifts. Since $\Lambda CDM$ is in fine agreement with empirical data, we set it as a benchmark for comparison in each analysis. The common redshift range is considered $z \in [0,1]$, where we observe that the models agree with the cosmographic parameters as well as the transition redshifts. 
Since cosmography confines only cosmological quantities that are not strictly dependent on a model, it alleviates degeneracy. The utilization of Taylor series expansion highly permits this alleviation in the lower redshift range. However, for a higher redshift range, this may lead to some extent of degeneracy \cite{Capozziello:2017uam}.

\begin{figure}[]
    \centering
    \includegraphics[width=0.7\linewidth]{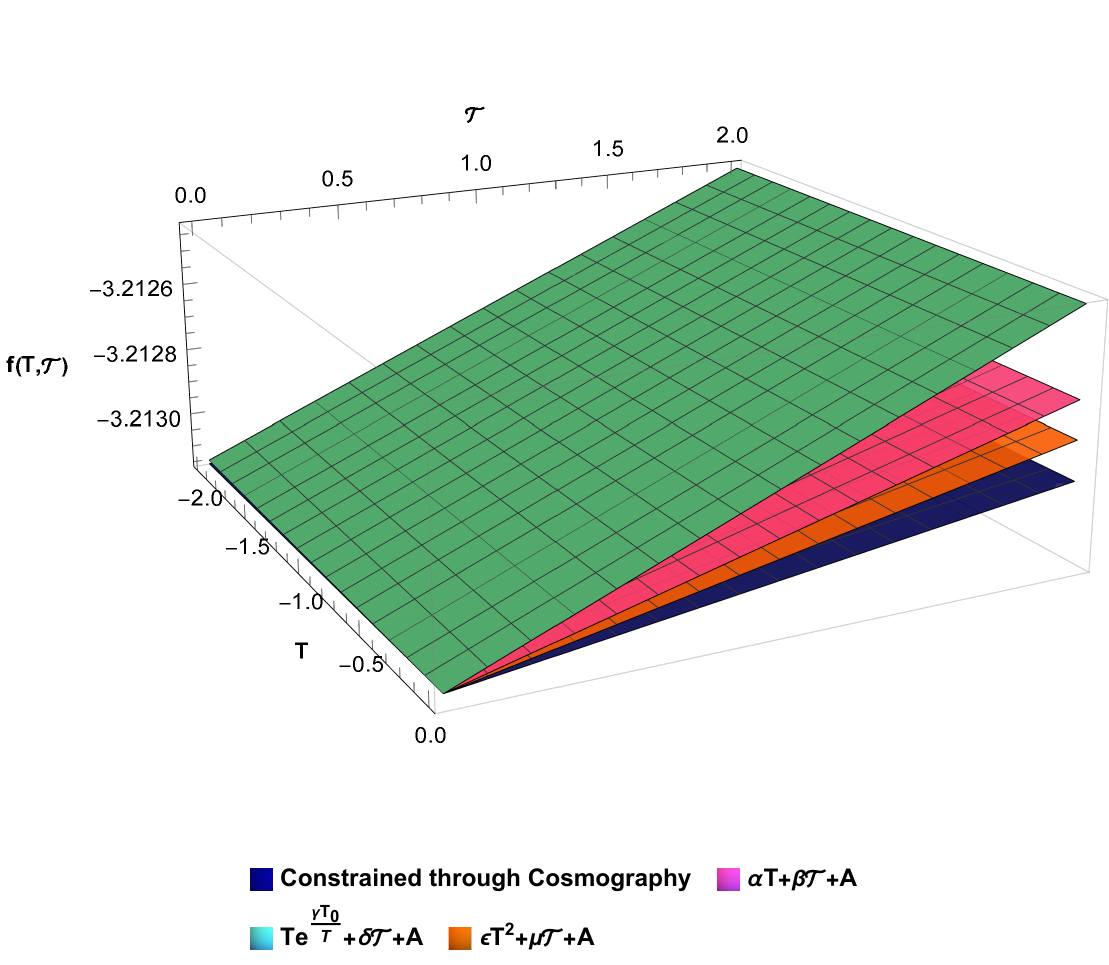}
    \caption{Comparision of three $f(T,\mathcal{T})$ models with the constrained function. The considered values of model parameters are $\alpha=3.5,\,\ \beta=2.25 \times 10^{14},\,\ \gamma=-0.00336,\,\ \delta =3.55 \times 10^{14}, \,\ A =-3.21314\times 10^{18}, \,\ \mu=1.7 \times 10^{14} \,\ \text{and} \,\ \epsilon=0.75.$}
    \label{fun}
\end{figure}

As the work has been executed in a model-independent manner, we interpret that any minimally coupled class of model from $f(T,\mathcal{T})$ theory should show a similar kind of behavior to the reconstructed functional form. For instance, in \autoref{fun} we consider the well-known exponential model $\bf{f(T,\mathcal{T})=T e^{\frac{\gamma T_0}{T}}+\delta \mathcal{T}+A}$, the linear model $\bf{f(T,\mathcal{T})=\alpha T +\beta \mathcal{T}+A}$ and the quadratic correction $\bf{f(T,\mathcal{T})=\epsilon T^2 +\mu \mathcal{T}+A}$ to compare with the constrained model. From the literature review, the above particular models have been taken into account due to their efficiency in explaining the evolution of the Universe. For certain fixed values of the free parameters, we find that all three models make an excellent match with our model (free parameter values mentioned in Fig. Caption). While explaining the dark sector responsible for the evolution of the universe, scalar-tensor theories provide promising results, as one can see in \cite{Brans:1961sx, Fujii:2003pa, Faraoni:2004pi}. Since the value of $\alpha$ shows a deviation from the usual value steaming from the gravitational constant when matching with the model-independent Taylor series form, we believe there is potential in this approach to connect the results of scalar-tensor theories. However, to make a definitive claim, we need to investigate its behavior in the early stages of the universe. Given that the Taylor series works well in regions with lower redshifts, we could obtain interesting results if the study is conducted using different orthogonal polynomials or series expansion with the intention of exploring the scenarios at higher redshifts. If this approach sheds light on scalar-tensor theories through cosmography, it could lead to revolutionary results. Since this is a promising idea, one can aim to reconnect scalar-tensor theory with cosmography using different series expansions as a future perspective.

One can also consider the non-minimally coupled form to constrain the theory using this method, such as $\bf{g(T) \times h(\mathcal{T})}$, $\bf{exp(g(T)+ h(\mathcal{T}))}$ etc. In order to cosmographically connect the early Universe with its late-time dynamics, the analysis can be extended to higher redshift using the other series expansions such as Pade polynomials, and Chebyshev polynomials. If one can successfully establish this connection, then the dark energy candidate can be fully explained solely in terms of cosmographic parameters, without necessarily applying prior modifications to Einstein's equations. This could potentially open up new dimensions of research.\\
 
\textbf{Data availability:} There are no new data associated with this article.\\

\begin{acknowledgments}  SSM acknowledges the Council of Scientific and Industrial Research (CSIR), Govt. of India for awarding Junior Research fellowship (E-Certificate No.: JUN21C05815). PKS acknowledges Science and Engineering Research Board, Department of Science and Technology, Government of India for financial support to carry out Research project No.: CRG/2022/001847. NSK, and VV acknowledge DST, New Delhi, India, for its financial support for research facilities under DST-FIST-2019. We are very much grateful to the honorable referee and to the editor for the illuminating suggestions that have significantly improved our work in terms of research quality, and presentation.
\end{acknowledgments}
\bibliography{apj}

\begin{thebibliography}{}
\expandafter\ifx\csname natexlab\endcsname\relax\def\natexlab#1{#1}\fi
\providecommand{\url}[1]{\href{#1}{#1}}
\providecommand{\dodoi}[1]{doi:~\href{http://doi.org/#1}{\nolinkurl{#1}}}
\providecommand{\doeprint}[1]{\href{http://ascl.net/#1}{\nolinkurl{http://ascl.net/#1}}}
\providecommand{\doarXiv}[1]{\href{https://arxiv.org/abs/#1}{\nolinkurl{https://arxiv.org/abs/#1}}}

\bibitem[{Abdalla {et~al.}(2022)}]{Abdalla:2022yfr}
Abdalla, E., {et~al.} 2022, JHEAp, 34, 49, \dodoi{10.1016/j.jheap.2022.04.002}

\bibitem[{Anderson {et~al.}(2014)}]{BOSS:2013rlg}
Anderson, L., {et~al.} 2014, Mon. Not. Roy. Astron. Soc., 441, 24, \dodoi{10.1093/mnras/stu523}

\bibitem[{Arouko {et~al.}(2020)Arouko, Salako, Kanfon, Houndjo, \& Baffou}]{Arouko:2020gbi}
Arouko, M.~Z., Salako, I.~G., Kanfon, A.~D., Houndjo, M. J.~S., \& Baffou, E. 2020, Int. J. Geom. Meth. Mod. Phys., 17, 2050116, \dodoi{10.1142/S0219887820501169}

\bibitem[{Bahamonde {et~al.}(2022)Bahamonde, Golovnev, Guzm\'an, Said, \& Pfeifer}]{Bahamonde:2021srr}
Bahamonde, S., Golovnev, A., Guzm\'an, M.-J., Said, J.~L., \& Pfeifer, C. 2022, JCAP, 01, 037, \dodoi{10.1088/1475-7516/2022/01/037}

\bibitem[{Bautista {et~al.}(2017)}]{Bautista:2017zgn}
Bautista, J.~E., {et~al.} 2017, Astron. Astrophys., 603, A12, \dodoi{10.1051/0004-6361/201730533}

\bibitem[{Blake {et~al.}(2012)}]{Blake:2012pj}
Blake, C., {et~al.} 2012, Mon. Not. Roy. Astron. Soc., 425, 405, \dodoi{10.1111/j.1365-2966.2012.21473.x}

\bibitem[{Brans \& Dicke(1961)}]{Brans:1961sx}
Brans, C., \& Dicke, R.~H. 1961, Phys. Rev., 124, 925, \dodoi{10.1103/PhysRev.124.925}

\bibitem[{Brout {et~al.}(2022{\natexlab{a}})}]{Brout:2022vxf}
Brout, D., {et~al.} 2022{\natexlab{a}}, Astrophys. J., 938, 110, \dodoi{10.3847/1538-4357/ac8e04}

\bibitem[{Brout {et~al.}(2022{\natexlab{b}})}]{Brout:2021mpj}
---. 2022{\natexlab{b}}, Astrophys. J., 938, 111, \dodoi{10.3847/1538-4357/ac8bcc}

\bibitem[{Burgess(2015)}]{Burgess:2013ara}
Burgess, C.~P. 2015, in {100e Ecole d'Ete de Physique: Post-Planck Cosmology}, 149--197, \dodoi{10.1093/acprof:oso/9780198728856.003.0004}

\bibitem[{Busca {et~al.}(2013)}]{Busca:2012bu}
Busca, N.~G., {et~al.} 2013, Astron. Astrophys., 552, A96, \dodoi{10.1051/0004-6361/201220724}

\bibitem[{Busti {et~al.}(2015)Busti, de~la Cruz-Dombriz, Dunsby, \& S\'aez-G\'omez}]{Busti:2015xqa}
Busti, V.~C., de~la Cruz-Dombriz, A., Dunsby, P. K.~S., \& S\'aez-G\'omez, D. 2015, Phys. Rev. D, 92, 123512, \dodoi{10.1103/PhysRevD.92.123512}

\bibitem[{Cai {et~al.}(2016)Cai, Capozziello, De~Laurentis, \& Saridakis}]{Cai:2015emx}
Cai, Y.-F., Capozziello, S., De~Laurentis, M., \& Saridakis, E.~N. 2016, Rept. Prog. Phys., 79, 106901, \dodoi{10.1088/0034-4885/79/10/106901}

\bibitem[{Capozziello {et~al.}(2011)Capozziello, Cardone, Farajollahi, \& Ravanpak}]{Capozziello:2011hj}
Capozziello, S., Cardone, V.~F., Farajollahi, H., \& Ravanpak, A. 2011, Phys. Rev. D, 84, 043527, \dodoi{10.1103/PhysRevD.84.043527}

\bibitem[{Capozziello {et~al.}(2008)Capozziello, Cardone, \& Salzano}]{Capozziello:2008qc}
Capozziello, S., Cardone, V.~F., \& Salzano, V. 2008, Phys. Rev. D, 78, 063504, \dodoi{10.1103/PhysRevD.78.063504}

\bibitem[{Capozziello {et~al.}(2017)Capozziello, D'Agostino, \& Luongo}]{Capozziello:2017uam}
Capozziello, S., D'Agostino, R., \& Luongo, O. 2017, Gen. Rel. Grav., 49, 141, \dodoi{10.1007/s10714-017-2304-x}

\bibitem[{Chen {et~al.}(2024)Chen, Wang, Zu, \& Fan}]{Chen:2024xkv}
Chen, R., Wang, Y.-Y., Zu, L., \& Fan, Y.-Z. 2024, Phys. Rev. D, 109, 024041, \dodoi{10.1103/PhysRevD.109.024041}

\bibitem[{Chen {et~al.}(2011)Chen, Dent, Dutta, \& Saridakis}]{Chen:2010va}
Chen, S.-H., Dent, J.~B., Dutta, S., \& Saridakis, E.~N. 2011, Phys. Rev. D, 83, 023508, \dodoi{10.1103/PhysRevD.83.023508}

\bibitem[{Choudhury {et~al.}(2019)Choudhury, Dasgupta, \& Banerjee}]{Choudhury:2019zod}
Choudhury, S.~G., Dasgupta, A., \& Banerjee, N. 2019, Mon. Not. Roy. Astron. Soc., 485, 5693, \dodoi{10.1093/mnras/stz731}

\bibitem[{Chuang \& Wang(2013)}]{Chuang:2012qt}
Chuang, C.-H., \& Wang, Y. 2013, Mon. Not. Roy. Astron. Soc., 435, 255, \dodoi{10.1093/mnras/stt1290}

\bibitem[{Chuang {et~al.}(2013)}]{Chuang:2013hya}
Chuang, C.-H., {et~al.} 2013, Mon. Not. Roy. Astron. Soc., 433, 3559, \dodoi{10.1093/mnras/stt988}

\bibitem[{Delubac {et~al.}(2015)}]{BOSS:2014hwf}
Delubac, T., {et~al.} 2015, Astron. Astrophys., 574, A59, \dodoi{10.1051/0004-6361/201423969}

\bibitem[{Demianski {et~al.}(2012)Demianski, Piedipalumbo, Rubano, \& Scudellaro}]{Demianski:2012ra}
Demianski, M., Piedipalumbo, E., Rubano, C., \& Scudellaro, P. 2012, Mon. Not. Roy. Astron. Soc., 426, 1396, \dodoi{10.1111/j.1365-2966.2012.21568.x}

\bibitem[{Dent {et~al.}(2011)Dent, Dutta, \& Saridakis}]{Dent:2010nbw}
Dent, J.~B., Dutta, S., \& Saridakis, E.~N. 2011, JCAP, 01, 009, \dodoi{10.1088/1475-7516/2011/01/009}

\bibitem[{Escamilla-Rivera \& Levi~Said(2020)}]{Escamilla-Rivera:2019ulu}
Escamilla-Rivera, C., \& Levi~Said, J. 2020, Class. Quant. Grav., 37, 165002, \dodoi{10.1088/1361-6382/ab939c}

\bibitem[{Faraoni(2004)}]{Faraoni:2004pi}
Faraoni, V. 2004, {Cosmology in scalar tensor gravity}, \dodoi{10.1007/978-1-4020-1989-0}

\bibitem[{Farrugia \& Levi~Said(2016)}]{Farrugia:2016pjh}
Farrugia, G., \& Levi~Said, J. 2016, Phys. Rev. D, 94, 124004, \dodoi{10.1103/PhysRevD.94.124004}

\bibitem[{Ferraro \& Fiorini(2007)}]{Ferraro:2006jd}
Ferraro, R., \& Fiorini, F. 2007, Phys. Rev. D, 75, 084031, \dodoi{10.1103/PhysRevD.75.084031}

\bibitem[{Font-Ribera {et~al.}(2014)}]{BOSS:2013igd}
Font-Ribera, A., {et~al.} 2014, JCAP, 05, 027, \dodoi{10.1088/1475-7516/2014/05/027}

\bibitem[{Fujii \& Maeda(2007)}]{Fujii:2003pa}
Fujii, Y., \& Maeda, K. 2007, {The scalar-tensor theory of gravitation}, Cambridge Monographs on Mathematical Physics (Cambridge University Press), \dodoi{10.1017/CBO9780511535093}

\bibitem[{Gadbail {et~al.}(2024)Gadbail, Arora, Sahoo, \& Bamba}]{Gadbail:2024syv}
Gadbail, G.~N., Arora, S., Sahoo, P.~K., \& Bamba, K. 2024.
\newblock \doarXiv{2402.04813}

\bibitem[{Ganiou {et~al.}(2016)Ganiou, Salako, Houndjo, \& Tossa}]{Ganiou:2015lvv}
Ganiou, M.~G., Salako, I.~G., Houndjo, M. J.~S., \& Tossa, J. 2016, Astrophys. Space Sci., 361, 57, \dodoi{10.1007/s10509-015-2644-5}

\bibitem[{Gaztanaga {et~al.}(2009)Gaztanaga, Cabre, \& Hui}]{Gaztanaga:2008xz}
Gaztanaga, E., Cabre, A., \& Hui, L. 2009, Mon. Not. Roy. Astron. Soc., 399, 1663, \dodoi{10.1111/j.1365-2966.2009.15405.x}

\bibitem[{Ghorui {et~al.}(2023)Ghorui, Rudra, \& Rahaman}]{Ghorui:2023ssn}
Ghorui, T., Rudra, P., \& Rahaman, F. 2023, Phys. Dark Univ., 42, 101352, \dodoi{10.1016/j.dark.2023.101352}

\bibitem[{Gomes {et~al.}(2024)Gomes, Briffa, Kozak, Levi~Said, Saal, \& Wojnar}]{Gomes:2023xzk}
Gomes, D.~A., Briffa, R., Kozak, A., {et~al.} 2024, JCAP, 01, 011, \dodoi{10.1088/1475-7516/2024/01/011}

\bibitem[{Gonzalez-Espinoza {et~al.}(2021)Gonzalez-Espinoza, Otalora, \& Saavedra}]{Gonzalez-Espinoza:2021mwr}
Gonzalez-Espinoza, M., Otalora, G., \& Saavedra, J. 2021, JCAP, 10, 007, \dodoi{10.1088/1475-7516/2021/10/007}

\bibitem[{Harko {et~al.}(2014)Harko, Lobo, Otalora, \& Saridakis}]{Harko:2014aja}
Harko, T., Lobo, F. S.~N., Otalora, G., \& Saridakis, E.~N. 2014, JCAP, 12, 021, \dodoi{10.1088/1475-7516/2014/12/021}

\bibitem[{Jiaze {et~al.}(2024)Jiaze, Zhihuan, Minghui, Rui, Jianping, \& Lixin}]{Gao:2024}
Jiaze, G., Zhihuan, Z., Minghui, D., {et~al.} 2024, Mon. Not. Roy. Astron. Soc., 527, 7861

\bibitem[{Jimenez {et~al.}(2003)Jimenez, Verde, Treu, \& Stern}]{Jimenez:2003iv}
Jimenez, R., Verde, L., Treu, T., \& Stern, D. 2003, Astrophys. J., 593, 622, \dodoi{10.1086/376595}

\bibitem[{Junior {et~al.}(2016)Junior, Rodrigues, Salako, \& Houndjo}]{Junior:2015bva}
Junior, E. L.~B., Rodrigues, M.~E., Salako, I.~G., \& Houndjo, M. J.~S. 2016, Class. Quant. Grav., 33, 125006, \dodoi{10.1088/0264-9381/33/12/125006}

\bibitem[{Kofinas \& Saridakis(2014)}]{Kofinas:2014daa}
Kofinas, G., \& Saridakis, E.~N. 2014, Phys. Rev. D, 90, 084045, \dodoi{10.1103/PhysRevD.90.084045}

\bibitem[{Linder(2010)}]{Linder:2010py}
Linder, E.~V. 2010, Phys. Rev. D, 81, 127301, \dodoi{10.1103/PhysRevD.81.127301}

\bibitem[{Luongo(2013)}]{Luongo:2013rba}
Luongo, O. 2013, Mod. Phys. Lett. A, 28, 1350080, \dodoi{10.1142/S0217732313500806}

\bibitem[{Malekjani {et~al.}(2023)Malekjani, Conville, Colg\'ain, Pourojaghi, \& Sheikh-Jabbari}]{Malekjani:2023dky}
Malekjani, M., Conville, R.~M., Colg\'ain, E.~O., Pourojaghi, S., \& Sheikh-Jabbari, M.~M. 2023.
\newblock \doarXiv{2301.12725}

\bibitem[{Maluf {et~al.}(2002)Maluf, da~Rocha-Neto, Toribio, \& Castello-Branco}]{Maluf:2002zc}
Maluf, J.~W., da~Rocha-Neto, J.~F., Toribio, T. M.~L., \& Castello-Branco, K.~H. 2002, Phys. Rev. D, 65, 124001, \dodoi{10.1103/PhysRevD.65.124001}

\bibitem[{Mandal {et~al.}(2023)Mandal, Sokoliuk, Mishra, \& Sahoo}]{Mandal:2023bzo}
Mandal, S., Sokoliuk, O., Mishra, S.~S., \& Sahoo, P.~K. 2023, Nucl. Phys. B, 993, 116285, \dodoi{10.1016/j.nuclphysb.2023.116285}

\bibitem[{Mandal {et~al.}(2020)Mandal, Wang, \& Sahoo}]{Mandal:2020buf}
Mandal, S., Wang, D., \& Sahoo, P.~K. 2020, Phys. Rev. D, 102, 124029, \dodoi{10.1103/PhysRevD.102.124029}

\bibitem[{Mishra {et~al.}(2024)Mishra, Kolhatkar, \& Sahoo}]{Mishra:2023onl}
Mishra, S.~S., Kolhatkar, A., \& Sahoo, P.~K. 2024, Phys. Lett. B, 848, 138391, \dodoi{10.1016/j.physletb.2023.138391}

\bibitem[{Mishra {et~al.}(2023)Mishra, Mandal, \& Sahoo}]{Mishra:2023khd}
Mishra, S.~S., Mandal, S., \& Sahoo, P.~K. 2023, Phys. Lett. B, 842, 137959, \dodoi{10.1016/j.physletb.2023.137959}

\bibitem[{Moreira {et~al.}(2021{\natexlab{a}})Moreira, Lima, Silva, \& Almeida}]{Moreira:2021uod}
Moreira, A. R.~P., Lima, F. C.~E., Silva, J. E.~G., \& Almeida, C. A.~S. 2021{\natexlab{a}}, Eur. Phys. J. C, 81, 1081, \dodoi{10.1140/epjc/s10052-021-09883-2}

\bibitem[{Moreira {et~al.}(2021{\natexlab{b}})Moreira, Silva, Lima, \& Almeida}]{Moreira:2021xfe}
Moreira, A. R.~P., Silva, J. E.~G., Lima, F. C.~E., \& Almeida, C. A.~S. 2021{\natexlab{b}}, Phys. Rev. D, 103, 064046, \dodoi{10.1103/PhysRevD.103.064046}

\bibitem[{Moresco(2015)}]{Moresco:2015cya}
Moresco, M. 2015, Mon. Not. Roy. Astron. Soc., 450, L16, \dodoi{10.1093/mnrasl/slv037}

\bibitem[{Moresco {et~al.}(2012)}]{Moresco:2012jh}
Moresco, M., {et~al.} 2012, JCAP, 08, 006, \dodoi{10.1088/1475-7516/2012/08/006}

\bibitem[{Moresco {et~al.}(2016)Moresco, Pozzetti, Cimatti, Jimenez, Maraston, Verde, Thomas, Citro, Tojeiro, \& Wilkinson}]{Moresco:2016mzx}
Moresco, M., Pozzetti, L., Cimatti, A., {et~al.} 2016, JCAP, 05, 014, \dodoi{10.1088/1475-7516/2016/05/014}

\bibitem[{Muthukrishna \& Parkinson(2016)}]{Muthukrishna:2016evq}
Muthukrishna, D., \& Parkinson, D. 2016, JCAP, 11, 052, \dodoi{10.1088/1475-7516/2016/11/052}

\bibitem[{Obukhov \& Pereira(2003)}]{Obukhov:2002tm}
Obukhov, Y.~N., \& Pereira, J.~G. 2003, Phys. Rev. D, 67, 044016, \dodoi{10.1103/PhysRevD.67.044016}

\bibitem[{Oka {et~al.}(2014)Oka, Saito, Nishimichi, Taruya, \& Yamamoto}]{Oka:2013cba}
Oka, A., Saito, S., Nishimichi, T., Taruya, A., \& Yamamoto, K. 2014, Mon. Not. Roy. Astron. Soc., 439, 2515, \dodoi{10.1093/mnras/stu111}

\bibitem[{Pace \& Said(2017{\natexlab{a}})}]{Pace:2017aon}
Pace, M., \& Said, J.~L. 2017{\natexlab{a}}, Eur. Phys. J. C, 77, 62, \dodoi{10.1140/epjc/s10052-017-4637-8}

\bibitem[{Pace \& Said(2017{\natexlab{b}})}]{Pace:2017dpu}
---. 2017{\natexlab{b}}, Eur. Phys. J. C, 77, 283, \dodoi{10.1140/epjc/s10052-017-4838-1}

\bibitem[{Panda {et~al.}(2024)Panda, Das, Manna, \& Ray}]{Panda:2023jwe}
Panda, A., Das, S., Manna, G., \& Ray, S. 2024, Phys. Dark Univ., 43, 101397, \dodoi{10.1016/j.dark.2023.101397}

\bibitem[{Perivolaropoulos \& Skara(2023)}]{Perivolaropoulos:2023iqj}
Perivolaropoulos, L., \& Skara, F. 2023, Mon. Not. Roy. Astron. Soc., 520, 5110, \dodoi{10.1093/mnras/stad451}

\bibitem[{Perlmutter {et~al.}(1997)}]{SupernovaCosmologyProject:1997czu}
Perlmutter, S., {et~al.} 1997, Bull. Am. Astron. Soc., 29, 1351.
\newblock \doarXiv{astro-ph/9812473}

\bibitem[{Piedipalumbo {et~al.}(2015)Piedipalumbo, Della~Moglie, \& Cianci}]{Piedipalumbo:2015jya}
Piedipalumbo, E., Della~Moglie, E., \& Cianci, R. 2015, Int. J. Mod. Phys. D, 24, 1550100, \dodoi{10.1142/S021827181550100X}

\bibitem[{Poplawski(2006)}]{Poplawski:2006na}
Poplawski, N.~J. 2006, Phys. Lett. B, 640, 135, \dodoi{10.1016/j.physletb.2006.07.056}

\bibitem[{Poplawski(2007)}]{Poplawski:2006ew}
---. 2007, Class. Quant. Grav., 24, 3013, \dodoi{10.1088/0264-9381/24/11/014}

\bibitem[{Ratsimbazafy {et~al.}(2017)Ratsimbazafy, Loubser, Crawford, Cress, Bassett, Nichol, \& V\"ais\"anen}]{Ratsimbazafy:2017vga}
Ratsimbazafy, A.~L., Loubser, S.~I., Crawford, S.~M., {et~al.} 2017, Mon. Not. Roy. Astron. Soc., 467, 3239, \dodoi{10.1093/mnras/stx301}

\bibitem[{Riess {et~al.}(1998)}]{SupernovaSearchTeam:1998fmf}
Riess, A.~G., {et~al.} 1998, Astron. J., 116, 1009, \dodoi{10.1086/300499}

\bibitem[{Riess {et~al.}(2022)}]{Riess:2021jrx}
---. 2022, Astrophys. J. Lett., 934, L7, \dodoi{10.3847/2041-8213/ac5c5b}

\bibitem[{Ross {et~al.}(2015)Ross, Samushia, Howlett, Percival, Burden, \& Manera}]{Ross:2014qpa}
Ross, A.~J., Samushia, L., Howlett, C., {et~al.} 2015, Mon. Not. Roy. Astron. Soc., 449, 835, \dodoi{10.1093/mnras/stv154}

\bibitem[{Sabiee {et~al.}(2022)Sabiee, Malekjani, \& Mohammad Zadeh~Jassur}]{Sabiee:2022iyo}
Sabiee, M., Malekjani, M., \& Mohammad Zadeh~Jassur, D. 2022, Mon. Not. Roy. Astron. Soc., 516, 2597, \dodoi{10.1093/mnras/stac2367}

\bibitem[{Saez-Gomez {et~al.}(2016)Saez-Gomez, Carvalho, Lobo, \& Tereno}]{Saez-Gomez:2016wxb}
Saez-Gomez, D., Carvalho, C.~S., Lobo, F. S.~N., \& Tereno, I. 2016, Phys. Rev. D, 94, 024034, \dodoi{10.1103/PhysRevD.94.024034}

\bibitem[{Scolnic {et~al.}(2022)}]{Scolnic:2021amr}
Scolnic, D., {et~al.} 2022, Astrophys. J., 938, 113, \dodoi{10.3847/1538-4357/ac8b7a}

\bibitem[{Simon {et~al.}(2005)Simon, Verde, \& Jimenez}]{Simon:2004tf}
Simon, J., Verde, L., \& Jimenez, R. 2005, Phys. Rev. D, 71, 123001, \dodoi{10.1103/PhysRevD.71.123001}

\bibitem[{Steinhardt {et~al.}(1999)Steinhardt, Wang, \& Zlatev}]{Steinhardt:1999nw}
Steinhardt, P.~J., Wang, L.-M., \& Zlatev, I. 1999, Phys. Rev. D, 59, 123504, \dodoi{10.1103/PhysRevD.59.123504}

\bibitem[{Stern {et~al.}(2010)Stern, Jimenez, Verde, Kamionkowski, \& Stanford}]{Stern:2009ep}
Stern, D., Jimenez, R., Verde, L., Kamionkowski, M., \& Stanford, S.~A. 2010, JCAP, 02, 008, \dodoi{10.1088/1475-7516/2010/02/008}

\bibitem[{Tonghua {et~al.}(2023)Tonghua, Shuo, Shuai, Yuting, Chenfa, \& Jieci}]{Tonghua:2023hdz}
Tonghua, L., Shuo, C., Shuai, M., {et~al.} 2023, Phys. Lett. B, 838, 137687, \dodoi{10.1016/j.physletb.2023.137687}

\bibitem[{Tsujikawa(2010)}]{Tsujikawa:2010sc}
Tsujikawa, S. 2010, \dodoi{10.1007/978-90-481-8685-3\_8}

\bibitem[{Unzicker \& Case(2005)}]{Unzicker:2005in}
Unzicker, A., \& Case, T. 2005.
\newblock \doarXiv{physics/0503046}

\bibitem[{Visser(2005)}]{Visser:2004bf}
Visser, M. 2005, Gen. Rel. Grav., 37, 1541, \dodoi{10.1007/s10714-005-0134-8}

\bibitem[{Visser(2015)}]{Visser:2015iua}
---. 2015, Class. Quant. Grav., 32, 135007, \dodoi{10.1088/0264-9381/32/13/135007}

\bibitem[{Vogt {et~al.}(2024)Vogt, Bocquet, Davies, Mohr, \& Schmidt}]{Vogt:2024pws}
Vogt, S. M.~L., Bocquet, S., Davies, C.~T., Mohr, J.~J., \& Schmidt, F. 2024.
\newblock \doarXiv{2401.09959}

\bibitem[{Wang {et~al.}(2017)}]{BOSS:2016zkm}
Wang, Y., {et~al.} 2017, Mon. Not. Roy. Astron. Soc., 469, 3762, \dodoi{10.1093/mnras/stx1090}

\bibitem[{Weinberg(1989)}]{Weinberg:1988cp}
Weinberg, S. 1989, Rev. Mod. Phys., 61, 1, \dodoi{10.1103/RevModPhys.61.1}

\bibitem[{Wu \& Yu(2010)}]{Wu:2010mn}
Wu, P., \& Yu, H.~W. 2010, Phys. Lett. B, 693, 415, \dodoi{10.1016/j.physletb.2010.08.073}

\bibitem[{Xu \& Wang(2011)}]{Xu:2010hq}
Xu, L., \& Wang, Y. 2011, Phys. Lett. B, 702, 114, \dodoi{10.1016/j.physletb.2011.06.091}

\bibitem[{Zhang {et~al.}(2014)Zhang, Zhang, Yuan, Zhang, \& Sun}]{Zhang:2012mp}
Zhang, C., Zhang, H., Yuan, S., Zhang, T.-J., \& Sun, Y.-C. 2014, Res. Astron. Astrophys., 14, 1221, \dodoi{10.1088/1674-4527/14/10/002}

\bibitem[{Zlatev {et~al.}(1999)Zlatev, Wang, \& Steinhardt}]{Zlatev:1998tr}
Zlatev, I., Wang, L.-M., \& Steinhardt, P.~J. 1999, Phys. Rev. Lett., 82, 896, \dodoi{10.1103/PhysRevLett.82.896}

\end{thebibliography}
\bibliographystyle{aasjournal}
\nocite{}

%% This command is needed to show the entire author+affiliation list when
%% the collaboration and author truncation commands are used.  It has to
%% go at the end of the manuscript.
%\allauthors

%% Include this line if you are using the \added, \replaced, \deleted
%% commands to see a summary list of all changes at the end of the article.
%\listofchanges

\end{document}